\documentclass[aps,prb,reprint,superscriptaddress,longbibliography]{revtex4-2}
\usepackage{amssymb}
\usepackage[utf8]{inputenc}
\usepackage{graphicx}
\usepackage{bm,color,subfigure,amsmath}

\begin{document}

\title{Ultrafast perturbation of magnetic domains by optical pumping in a ferromagnetic multilayer}

\author{Dmitriy Zusin}
\affiliation{Department of Physics and JILA, University of Colorado, Boulder, CO 80309, USA}

\author{Ezio Iacocca}
\affiliation{Department of Applied Mathematics, University of Colorado, Boulder, CO 80309, USA}
\affiliation{Center for Magnetism and Magnetic Materials, University of Colorado Colorado Springs, Colorado Springs, CO 80918, USA}

\author{Lo\"{i}c Le Guyader}
\affiliation{Stanford Institute for Materials and Energy Sciences, SLAC National Accelerator Laboratory, 2575 Sand Hill Road, Menlo Park, CA 94025, USA}
\affiliation{European XFEL, Holzkoppel 4, 22869 Schenefeld, Germany}

\author{Alexander H. Reid}
\affiliation{Stanford Institute for Materials and Energy Sciences, SLAC National Accelerator Laboratory, 2575 Sand Hill Road, Menlo Park, CA 94025, USA}
\affiliation{Linac Coherent Light Source, SLAC National Accelerator Laboratory, 2575 Sand Hill Road, Menlo Park, CA 94025, USA}

\author{William F. Schlotter}
\affiliation{Linac Coherent Light Source, SLAC National Accelerator Laboratory, 2575 Sand Hill Road, Menlo Park, CA 94025, USA}

\author{Tian-Min Liu}
\affiliation{Stanford Institute for Materials and Energy Sciences, SLAC National Accelerator Laboratory, 2575 Sand Hill Road, Menlo Park, CA 94025, USA}

\author{Daniel J. Higley}
\affiliation{Stanford Institute for Materials and Energy Sciences, SLAC National Accelerator Laboratory, 2575 Sand Hill Road, Menlo Park, CA 94025, USA}

\author{Giacomo Coslovich}
\affiliation{Linac Coherent Light Source, SLAC National Accelerator Laboratory, 2575 Sand Hill Road, Menlo Park, CA 94025, USA}

\author{Scott F. Wandel}
\affiliation{Linac Coherent Light Source, SLAC National Accelerator Laboratory, 2575 Sand Hill Road, Menlo Park, CA 94025, USA}
\affiliation{Current address: MIT Lincoln Laboratory, 244 Wood Street, Lexington, MA 02421, USA}

\author{Phoebe M. Tengdin}
\affiliation{Department of Physics and JILA, University of Colorado, Boulder, CO 80309, USA}

\author{Sheena K. K. Patel}
\affiliation{Department of Physics, University of California San Diego, La Jolla, California 92093, USA}
\affiliation{Center for Memory and Recording Research, University of California San Diego, La Jolla, California 92093, USA}

\author{Anatoly Shabalin}
\affiliation{Department of Physics, University of California San Diego, La Jolla, California 92093, USA}

\author{Nelson Hua}
\affiliation{Department of Physics, University of California San Diego, La Jolla, California 92093, USA}
\affiliation{Center for Memory and Recording Research, University of California San Diego, La Jolla, California 92093, USA}

\author{Stjepan B. Hrkac}
\affiliation{Department of Physics, University of California San Diego, La Jolla, California 92093, USA}

\author{Hans T. Nembach}
\affiliation{Quantum Electromagnetics Division, National Institute of Standards and Technology, Boulder, Colorado 80305, USA}

\author{Justin M. Shaw}
\affiliation{Quantum Electromagnetics Division, National Institute of Standards and Technology, Boulder, Colorado 80305, USA}

\author{Sergio A. Montoya}
\affiliation{Center for Memory and Recording Research, University of California San Diego, La Jolla, California 92093, USA}

\author{Adam Blonsky}
\affiliation{Department of Physics and JILA, University of Colorado, Boulder, CO 80309, USA}

\author{Christian Gentry}
\affiliation{Department of Physics and JILA, University of Colorado, Boulder, CO 80309, USA}

\author{Mark A. Hoefer}
\affiliation{Department of Applied Mathematics, University of Colorado, Boulder, CO 80309, USA}

\author{Margaret M. Murnane}
\affiliation{Department of Physics and JILA, University of Colorado, Boulder, CO 80309, USA}

\author{Henry C. Kapteyn}
\affiliation{Department of Physics and JILA, University of Colorado, Boulder, CO 80309, USA}

\author{Eric E. Fullerton}
\affiliation{Center for Memory and Recording Research, University of California San Diego, La Jolla, California 92093, USA}

\author{Oleg Shpyrko}
\affiliation{Department of Physics, University of California San Diego, La Jolla, California 92093, USA}

\author{Hermann A. Dürr}
\affiliation{Department of Physics and Astronomy, Uppsala University, S-75120 Uppsala, Sweden}

\author{T. J. Silva}
\affiliation{Quantum Electromagnetics Division, National Institute of Standards and Technology, Boulder, Colorado 80305, USA}

\begin{abstract}
    Ultrafast optical pumping of spatially nonuniform magnetic textures is known to induce far-from-equilibrium spin transport effects. Here, we use ultrafast x-ray diffraction with unprecedented dynamic range to study the laser-induced dynamics of labyrinth domain networks in ferromagnetic CoFe/Ni multilayers. We detected azimuthally isotropic, odd order, magnetic diffraction rings up to 5th order. The amplitudes of all three diffraction rings quench to different degrees within 1.6 ps. In addition, all three of the detected diffraction rings both broaden by 15\% and radially contract by 6\% during the quench process. We are able to rigorously quantify a 31\% ultrafast broadening of the domain walls via Fourier analysis of the order-dependent quenching of the three detected diffraction rings. The broadening of the diffraction rings is interpreted as a reduction in the domain coherence length, but the shift in the ring radius, while unambiguous in its occurrence, remains unexplained. In particular, we demonstrate that a radial shift explained by domain wall broadening can be ruled out. With the unprecedented dynamic range of our data, our results provide convincing evidence that labyrinth domain structures are spatially perturbed at ultrafast speeds under far-from-equilibrium conditions, albeit the mechanism inducing the perturbations remains yet to be clarified.
\end{abstract}

\maketitle

\section{Introduction}

Understanding ultrafast magnetization processes~\cite{Beaurepaire1996,Koopmans2009,Ostler2012,Turgut_2013,Xu2017,Iimaha2018,Turgut_2016,Zusin2018,Eich2017,Tengdin2018} is challenging because of the strongly-coupled interactions between the charges, spins, and phonons. These are difficult to probe and model in equilibrium, and even more so when the material is subjected to a femtosecond laser pulse that floods the conduction band with a far-from-equilibrium, dense distribution of hot electrons. Many theoretical models and mechanisms have been proposed to explain experimental findings, including heat redistribution in the quasi-equilibrium spin, electronic, and lattice systems~\cite{Beaurepaire1996,Koopmans2009,Ostler2012} , superdiffusive spin currents into metallic spin sinks~\cite{Turgut_2013,Battiato2010}, Elliott-Yafet scattering~\cite{Turgut_2013}, hot-electron transport~\cite{Xu2017,Iimaha2018}, ultrafast magnon generation and exchange splitting reduction~\cite{Turgut_2016,LeGuyader2022}, and recently the observation of critical behavior and a magnetic phase transition within 20 fs~\cite{Eich2017,Tengdin2018}.

Many studies of ultrafast magnetization processes use spatially-averaged measurements such as x-ray magnetic circular dichroism (XMCD) or magneto-optic measurements using visible or x-ray light. More recently, it has been possible to study the impact of morphological and magnetic spatial inhomogeneities on the dynamic response of a material by taking advantage of the diffracted x-ray intensity to extract spatial information~\cite{Graves2013,Iacocca2019,Pfau2012,Vodungbo2012,ZhouHagstrom2022}. Transport mechanisms~\cite{Battiato2010,Rudolf2012,Eschenlohr2013} have been proposed to describe spatially-dependent ultrafast responses, such as the demagnetization and domain-wall broadening in domain networks~\cite{Pfau2012,Vodungbo2012,vonKorffSchmising2014,Hennes2020}, and the imprinting of domain patterns in ferrimagnetic metallic alloys~\cite{Graves2013,Iacocca2019}. An interesting signature of a dynamic nanoscopic spatial response is a time-dependent shift in the observed x-ray scattering. In Ref.~\onlinecite{Iacocca2019}, a ring contraction was associated with the transition from a morphologically-induced magnetisation pattern into nonlinear dynamical spin textures upon partial quenching of a homogeneously magnetised ferrimagnet. 

In Ref.~\onlinecite{Pfau2012}, a puzzling ultrafast shift of the first-order x-ray magnetic diffraction ring radius was first observed in the case of a labyrinth domain network. The authors hypothesized that the shift was a higher-order effect due to domain-wall broadening. Such broadening was predicted to occur as a result of superdiffusive spin current propagation across the domain walls~\cite{vonKorffSchmising2014}. However, the inability to detect any higher order diffraction rings prohibited quantitative testing of this hypothesis. Only a very weak or negligible shift in diffraction ring radius has been detected to date when the experiment is repeated with samples that exhibit stripe domain patterns stabilized by a weak external magnetic field ~\cite{Vodungbo2012,Hennes2020,ZhouHagstrom2022}. It was recently shown by use of samples supporting both stripe and labyrinth domain patterns that the shift in the diffraction ring radius occurs only with labyrinths and not with stripes, suggestive of a mechanism that is sensitive to domain symmetry~\cite{ZhouHagstrom2022}.

Here, we probe time-resolved x-ray diffraction from labyrinth domain networks in a CoFe/Ni multilayer with perpendicular magnetic anisotropy to discern the influence of domain-wall broadening on the shift of the diffraction rings. We are able to resolve up to the fifth-order diffraction ring with unprecedented dynamic range, enabling a quantitative determination of how ultrafast pumping affects both the domain-wall width and the magnetic correlation length. We rule out domain wall broadening as the cause of diffraction ring radius shift.

A 31\% ultrafast broadening of the domain-walls is rigorously quantified by fitting the relative quench of the first three diffraction ring amplitudes to a Bloch-wall model. In addition, we  detect a 15\% decrease in the domain correlation length –– from 845~nm $\pm$ 1~nm to 711~nm $\pm$ 2~nm within 1.6~ps. This surprising result is suggestive of an ultrafast spatial alteration of the domain structure, possibly the result of a zero-mean random domain wall displacements mediated by far-from-equilibrium electronic excitations. A 6\% contraction of the diffraction ring radii within 1.6~ps of laser excitation is simultaneously observed, confirming previous reports of such shifts. Because we can extract the true domain wall broadening via detection of the first three orders of diffraction rings, we conclusively exclude domain-wall broadening as the source of diffraction ring contraction, contrary to the original hypothesis in Ref.~\onlinecite{Pfau2012}. Our observation of significant distortions in the diffraction ring structure, which include amplitude, width, and radius, suggests that domain walls in labyrinth structures are to some extent mobile at ultrafast speeds when subjected to far-from-equilibrium conditions. It remains to be determined how such a surprising effect occurs. 


\section{Time-dependent x-ray scattering}
\begin{figure*}[t]
\centering \includegraphics[width=6.5in]{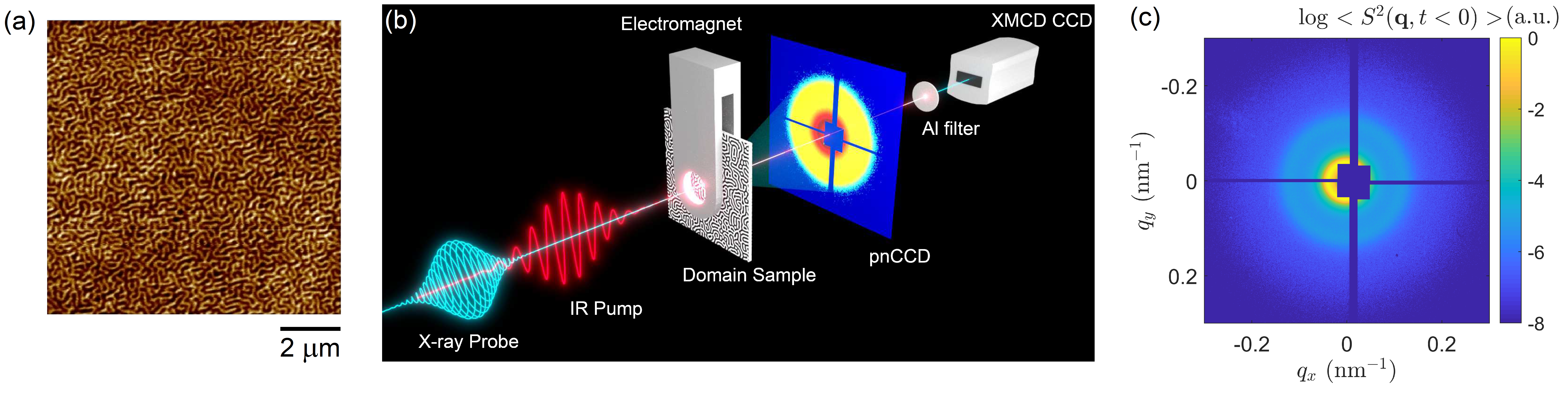}
\caption{ \label{fig1} (a) $10\times10$~$\mu$m$^2$ magnetic force microscope (MFM) image of a similar CoFe/Ni multilayer sample to that used for the x-ray scattering measurements. (b) An incident femtosecond infrared (IR) pulse excites the sample. The time-dependent magnetization is measured by a time-delayed, circularly-polarized x-ray probe. The scattered x-rays are captured by a primary, high-speed CCD while the unscattered beam is captured by a secondary CCD acting as a point detector. An electromagnet is used to saturate the sample, allowing for both measurements of time-resolved XMCD, as well as the static charge contribution to the scattered intensity. (c) Two-dimensional magnetic component of the scattered intensity obtained with the primary CCD. The first-order diffraction ring is partially obscured due to the aperture in the middle of the primary CCD. }
\end{figure*}
\begin{figure}[t]
\centering \includegraphics[width=3in]{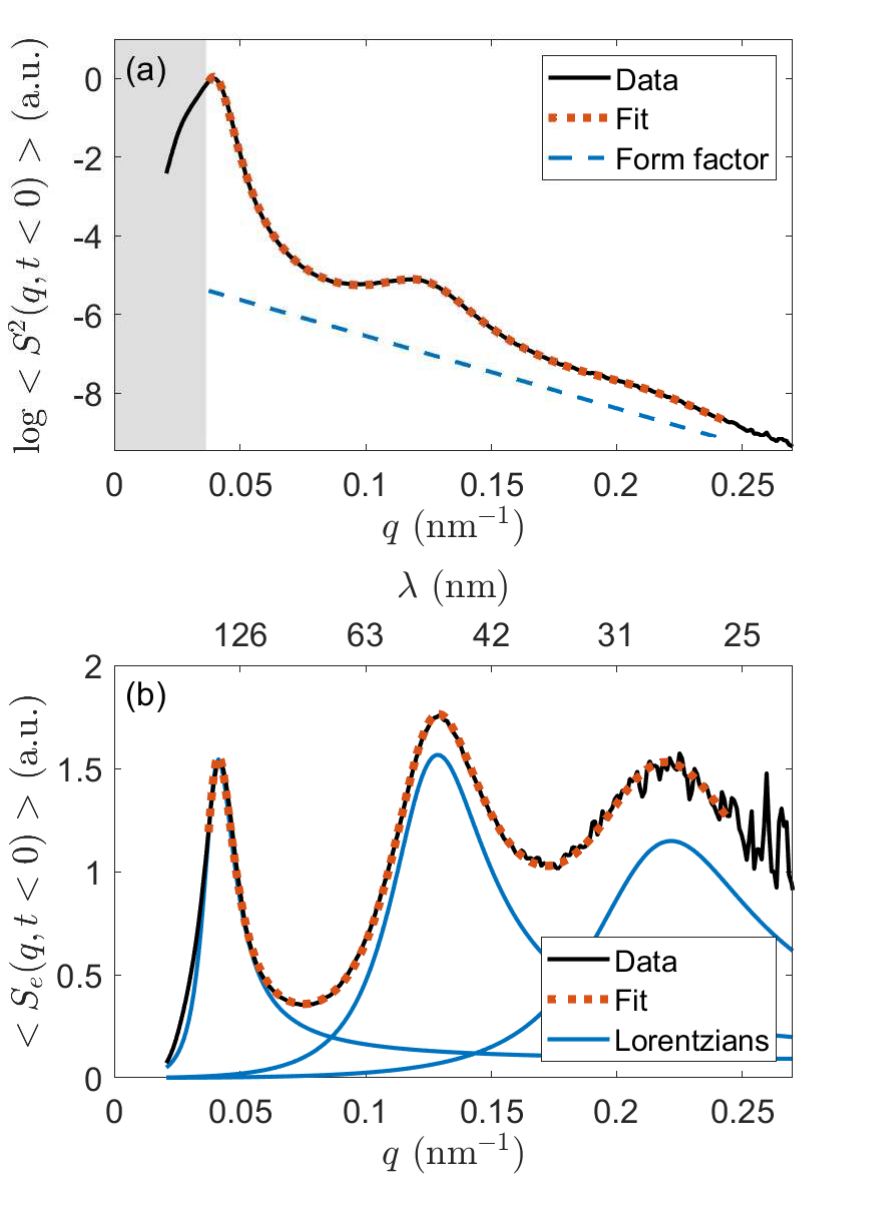}
\caption{ \label{fig2} (a) Equilibrium azimuthally-averaged magnetic scattering. The data, the fit to the data with Eq.~\eqref{eq:2}, and the fitted form factor are shown by the solid black, dashed red, and dashed blue curves, respectively. The same data and fit are shown in (b) after equalization, as per Eq.~\eqref{eq:3}, to accentuate the quality of the fit for all the diffraction rings. The solid blue curves represent the three Lorentzian components of the fit used to determine the periodic structure of the domains. The asymmetry of the Lorentzians is due to the power law scaling used to equalize the 1st and 5th order diffraction ring intensities. }
\end{figure}
We measured the picosecond time-evolution of the labyrinth domain network by use of pump-probe coherent, time-dependent, soft x-ray small-angle scattering at the Linac Coherent Light Source (LCLS) free-electron laser. The magnetic samples were fabricated by sputter deposition with the following layer composition: Si$_3$N$_4$(50) / Ta(3) / Cu(5) / [Co$_{90}$Fe$_{10}$(0.2)/ Ni(0.6)]x50 / CoFe(0.2) / Cu(3) / Ta(3), where the layer thicknesses in parentheses are in nm and the Si$_3$N$_4$ membrane allowed for x-ray transmission. The magnetic parameters of the 40~nm thick CoFe/Ni ferromagnetic multilayers were measured as a function of temperature with a vibrating sample magnetometer (VSM). At room temperature, we determine a saturation magnetization $M_s = 771$~kA/m, a first-order anisotropy constant $K_1 = 739$~kJ/m$^3$, and a negative second-order anisotropy constant $K_2 = -266$~kJ/m$^3$. A non-negligible second-order anisotropy was previously reported for this material system~\cite{Shaw2013}. The net uniaxial anisotropy, including the magnetostatic contribution, is $99$~kJ/m$^3$. This corresponds to an effective magnetization $M_\mathrm{eff} = -2.05$~kA/m for perpendicular ferromagnetic resonance (FMR). The experimentally measured FMR value is $M_\mathrm{eff} = -2.12$~kA/m, in good agreement with the VSM measured value. Despite the large second-order anisotropy constant, the relative magnitudes of the first- and second-order anisotropies are within the range necessary for a net perpendicular magnetic anisotropy~\cite{Hubert2009}. We confirmed that the out-of-plane labyrinth domain network is indeed stabilized at room temperature by use of magnetic force microscopy measurements with a spatial resolution of $\approx22$~nm, shown in Fig.~\ref{fig1}(a).

X-ray measurements were performed at the SXR hutch~\cite{Schlotter2012}. The experimental setup is schematically shown in Fig.~\ref{fig1}(b). The free-electron laser (FEL) generated 60~fs long soft x-ray pulses at a repetition rate of 120~Hz with a photon energy of 852.7~eV to match the L3 absorption edge of Ni. Circularly polarized x-rays were achieved by use of a Delta-undulator~\cite{Lutman2016}. The x-ray beam was focused to an elliptical spot with foci a = 23~$\mu$m and b = 15~$\mu$m. A high-speed primary pnCCD camera (Max Planck Semiconductor Laboratory supplied by PNSensor GmbH)  placed 275.3~mm away from the sample captured the time-dependent scalar diffracted intensity of the probe beam. The detector had four $512\times512$ pixel panels that could be moved independently from one another and each pixel had a maximum well-depth of 16,000 electrons. The CCD camera had an opening at the center through which unscattered x-rays were transmitted. These x-rays were detected with a secondary CCD camera (Andor Newton DO940P-BN) placed behind the primary CCD camera. An Al filter in front of the secondary CCD was used to suppress the infrared pump beam, which was collinear with the incident x-ray beam. In addition to scattering measurements without an applied magnetic field, we carried out measurements of both the scattering (with the primary CCD camera) and XMCD (with the secondary CCD camera) when the sample was magnetically saturated to remove any non-magnetic contributions from the zero-field scattering data~\cite{Highley2016} (see Appendix~\ref{app_a} for details). For this, an external magnetic field of $0.6$~T was applied perpendicularly to the surface of the sample. This experimental geometry allowed us to detect x-rays scattered at angles of up to $\approx8^\circ$.

An amplified infrared (IR) laser pump pulse from a Ti:Sapphire laser at the central wavelength of 795~nm was used to pump the sample. The duration of the IR pump pulses was 60~fs, the Gaussian beam waist size was 172~$\mu$m, and the average incident pump fluence was 23~mJ/cm$^{2}$. Higher pump fluence resulted in catastrophic damage to the sample. The pump laser was synchronized with the FEL to within the jitter of the arrival time of x-ray pulses. The delay time between the IR pump and the x-ray probe was varied from negative delays (to probe an unperturbed sample before the IR pump has arrived) to 20~ps. Scattering patterns at different delays were collected in a single-shot manner, and the pattern at a given delay was computed as an average of all of the scattering patterns taken within $\pm$200~fs of that delay.

The time-evolution of the labyrinth domain network is inferred from the squared magnetic scattering amplitude $|S(\mathbf{q},t)|^2$, with wavevector $\mathbf{q}$. We isolated this component from the diffracted intensity $I(\mathbf{q},t)$ by subtracting the charge intensity $|C(\mathbf{q},t)|^2$ obtained from the saturated sample, as described in Appendix~\ref{app_a}. For labyrinth domains randomly oriented in the film’s plane, $|S(\mathbf{q},t)|^2$ consists of concentric rings, shown in Fig.~\ref{fig1}(c). The first-order diffraction ring contained 500 electrons per shot, $\approx3$\% of the CCD camera saturation. We note that the first-order scattering ring is partially obscured by the location of the through-beam aperture in the center of the CCD camera, depicted as a dark-blue box in Fig.~\ref{fig1}(c).

We azimuthally average the magnetic scattering intensity to obtain $S^2(q,t)$, where $q=|\mathbf{q}|$. To account for the incomplete data captured by the primary CCD camera, we utilize the following algorithm. First, the center of the scattering pattern, $|\mathbf{q}|=0$, is determined by fitting a circle to the third-order diffraction ring. Because this diffraction ring was not obscured by the central square aperture, a reliable fit can be obtained for the center location in pixels. Once the center is determined, the data is then azimuthally averaged. By definition, the azimuthal average is
\begin{equation}
\label{eq:1}
    S^2(q,t)=\frac{\int_0^{2\pi}{S^2(q\cos{(\theta)},t)d\theta}}{L(q)},
\end{equation}
where $\varphi$ is the azimuth for the $\mathbf{q}$ vector and $L(q)$ is the circumference for a given $q$. To account for the missing pixels, we compute $L(q)=\int_0^{2\pi}{W(q,t)d\varphi}$, where $W(q,\varphi)$ is a two-dimensional mask of the CCD cameras and missing pixels are numerically counted as zeros. In this way, the azimuthal average is normalized by an adjusted circumference.

The pre-pump ($t < 0$) average data is shown in logarithmic scale in Fig.~\ref{fig2}(a) by a solid black curve. The shoulder in the first-order diffraction ring at $q < 0.0375$~nm$^{-1}$, shown by a grey area, is an artifact of the aforementioned partial obscuration by the aperture whereby the limited data is more sensitive to the number of pixels available when computing Eq.~\eqref{eq:1}. The fifth-order ring, as well as an exponentially decaying background, are clearly visible in the azimuthally averaged intensity.

\section{Empirical model and data fitting}

To extract information from the azimuthally averaged scattering $S^2(q,t)$, we make use of a Lorentzian empirical fitting function for the first three of the odd $n$th order diffraction rings
\begin{equation}
\label{eq:2}
    f(q,t) = e^{-2q/Q(t)}\left[M_0(t)+\sum_{n=1,3,5}{\frac{M_{n}(t)}{\left(\frac{q-nq_0(t)}{n\Gamma(t)}\right)^2+1}}\right]^2.
\end{equation}

The first factor outside of the square brackets is an exponential form factor we associate with the non-zero characteristic spin-spin correlation length scale, $Q(t)$. The term in the square brackets is the magnetic structure factor, consisting of a linear superposition of random uniform spatial fluctuations $M_0(t)$ and three Lorentzian diffraction rings centred at odd-integer multiples of the first-order ring position $q_0(t)$. $M_{n}(t)$ are the rings’ amplitudes with subscripts $n=1,3,5$ denoting the respective odd order diffraction ring, and the width of each diffraction ring or linewidth is parameterized by $\Gamma(t)$. Note that while each ring is fitted with an independent amplitude, the ring radii and widths are all constrained to be integer multiples of the diffraction order.

We stress that $f(q,t)$ is purely phenomenological; it was found by trial and error that application of such a function yields an excellent fit to the data. However, the applicability of a Lorentzial linewidth is consistent with an exponentially decaying autocorrelation function for the domain pattern. The fitting function proposed in Ref.~\onlinecite{Hellwig2003} was not used because the underlying model used in its derivation is only applicable for a system of parallel stripe domains with domain walls much narrower that the domain spacing.

The simultaneous fit of all three diffraction rings and the form factor allows us to to accurately determine all seven fitting parameters in $f(q)$. This approach takes advantage of all the available data and the integer multiple relationship between all three rings to obtain an unambiguous fit despite potential artifacts associated with the partial obscuration by the central aperture. The fits are performed on the logarithm of the scattering data to maximize sensitivity of the strongly attenuated 3rd and 5th order diffraction rings.

The fit of the time-averaged $t<0$ diffraction data by use of Eq.~\eqref{eq:2} is shown in Fig.~\ref{fig2}(a) by the red dashed curve. The fitted first-order ring radius is $q_0(0) = 0.0392$~nm$^{-1}\pm2\times10^{-5}$~nm$^{-1}$, equivalent to an equilibrium domain width of $\pi/q_0(0) = 80.1$~nm $\pm~0.01$~nm. Magnetic force microscopy imaging of the labyrinth domain network is comparable to this average domain width. The small error associated with the fitted parameters is a result of the simultaneous fitting of the harmonic third-order and fifth-order ring radii. 

The exponential form factor contribution of $Q(0) = 0.1087$~nm$^{-1}\pm4\times10^{-5}$~nm$^{-1}$ is shown by a dashed blue line. Because $Q$ corresponds to a spatial distribution of spin density with a Lorentzian-like correlation function, we may interpret it as an approximation of the exchange length, $\lambda_\mathrm{ex} \approx 1/Q = 9.19$~nm~$\pm~0.0038$~nm. This quantity is in rough agreement with the calculated exchange length of $7.3$~nm determined from a combination of magnetometry measurements and an assumed exchange constant of $A_\mathrm{ex}=20$~pJ/m, so that $\lambda_\mathrm{ex}=\sqrt{2A/(\mu_0M_s^2)}$.

To illustrate the quality of the fitting, we show the azimuthally averaged scattering in Fig.~\ref{fig1}(b) using an ad hoc equalized representation
\begin{equation}
\label{eq:3}
    S_e(q,t)=\left(\sqrt{S^2(q,t)}e^{(2q/Q)}-M_0(t)\right)q^{2.12},
\end{equation}
where the exponential form factor is divided out, the magnetic noise background $M_0(t)$ is subtracted, and an adjustable power law scale factor $q^{2.12}$ was chosen to equalize the amplitudes of the first-order and third-order rings. By use of this ad hoc equalization, the excellent fidelity of the fits is clearly apparent. The individual Lorentzian components are shown with solid blue curves. Again, we stress that the radii and widths of all three Lorentzians are constrained to be odd integer multiples of the 1st order diffraction ring.
\begin{figure*}[t]
\centering \includegraphics[width=6in]{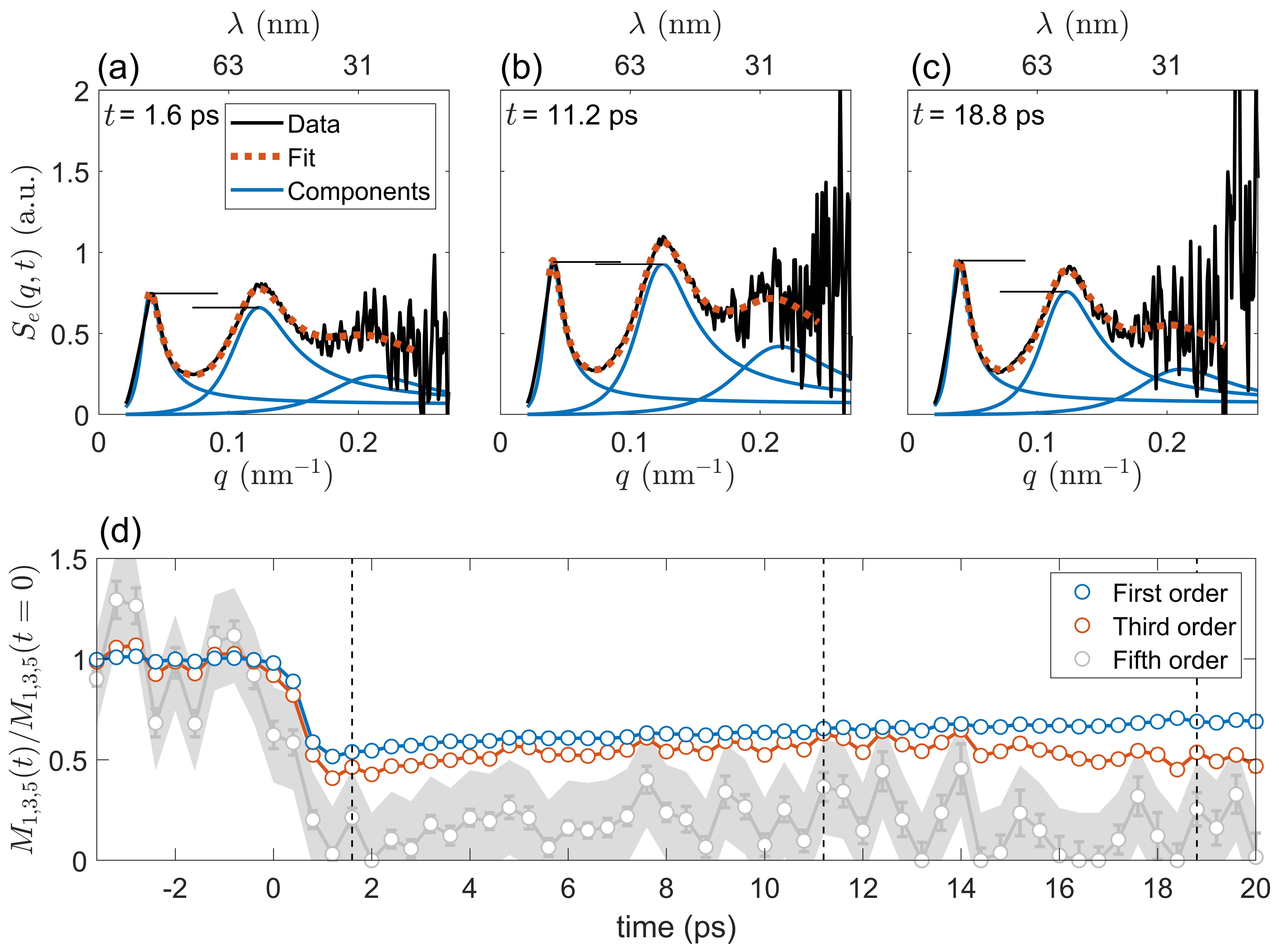}
\caption{ \label{fig3} The equalized data, fits, and Lorentzians, as per Eq.~\ref{eq:3}, are shown by solid black, dashed red, and solid blue curves, respectively for three instances in time after pumping: (a) 1.6 ps, (b) 11.2 ps, and (c) 18.8 ps. The horizontal black lines illustrate the relative amplitude of the first- and third-order diffraction rings. (d) Temporal amplitude evolution of the first- (blue), third- (red), and fifth-order (grey) diffraction rings. The vertical dashed lines correspond to the time instances shown in panels (a), (b), and (c). Error bars represent one standard deviation of the fitted quantities. The grey area represents the standard deviation of fifth-order magnitude by averaging the pre-pump ($t < 0$) data. }
\end{figure*}

\section{Fourier analysis of the equilibrium scattered spectra}

The empirical model of Eq.~\eqref{eq:2}, $f(q,t)$, provides accurate fits to the azimuthally averaged data, as shown in Fig.~\ref{fig2}. Therefore, we invoke concepts of Fourier analysis to interpret the salient features of the physical system captured by the functional form of $f(q,t$).

First, consider an ideal, 1D periodic function with period $x_0$. By Fourier series decomposition, its spectrum will be composed of harmonically related delta functions starting at the fundamental frequency $2\pi/x_0$. Such a spectrum is independent of the periodic function's profile or functional form. Instead, the profile is encoded in the relative amplitudes of the harmonic delta functions. In the case of a perfect sinusoidal function, the ratio is zero, meaning that only the fundamental harmonic exists. In the extreme case of a square wave, the ratio is $1/n$, with $n$ being the odd order index of the Fourier component. Any smooth profile will therefore exhibit components with amplitudes with an order dependence that varies between  $0$ (sine wave) and $1/n$ (square wave). The crucial statement here is that the spatial profile of the domain walls in a periodic lattice is principally encoded in the \textit{relative} amplitudes of the components, not in their widths. The diffraction ring widths are instead related to phase uncertainty for periodic structures.

Variations in the periodicity of a 1D oscillatory function, akin to jitter in temporal signals~\cite{Carlson2002}, introduces uncertainty in the component frequency. The greater the variations, the broader the individual Fourier components of the periodic function. Most importantly, the broadening scales with the integer order of the individual components, i.e. the fractional uncertainty in the periodicity of the domain structure is the same, regardless of the diffraction order of the ring. The form of Eq.~\eqref{eq:2} accounts for these fundamental properties of any periodic domain structure.

From the aforementioned properties of Fourier series, it becomes clear that the azimuthally averaged scattering provides two distinct types of information: 1) the position of the harmonic peaks is related to the average spatial frequency of the magnetic texture, and 2) the relative amplitude of the harmonic rings is related to the profile of the magnetic texture, i.e., the domain-wall width. These properties have profound implications in the interpretation of the time-dependent modifications of the scattering.

It is worth pointing out that this analysis is rooted in linear response theory. In other words, these arguments hold as long as the system does not exhibit nonlinearities. If nonlinearities are present, the Fourier spectrum can indeed exhibit an artificial shift based on the distortion of the underlying waveform. Such a shift, however, is accompanied by a spectral distortion of the peak itself. As we show below, our experiments exhibited no discernible spectral distortion, suggesting that linear response theory is appropriate.

\section{Ultrafast modification of the diffraction pattern}

\begin{figure}[t]
\centering \includegraphics[width=3.3in]{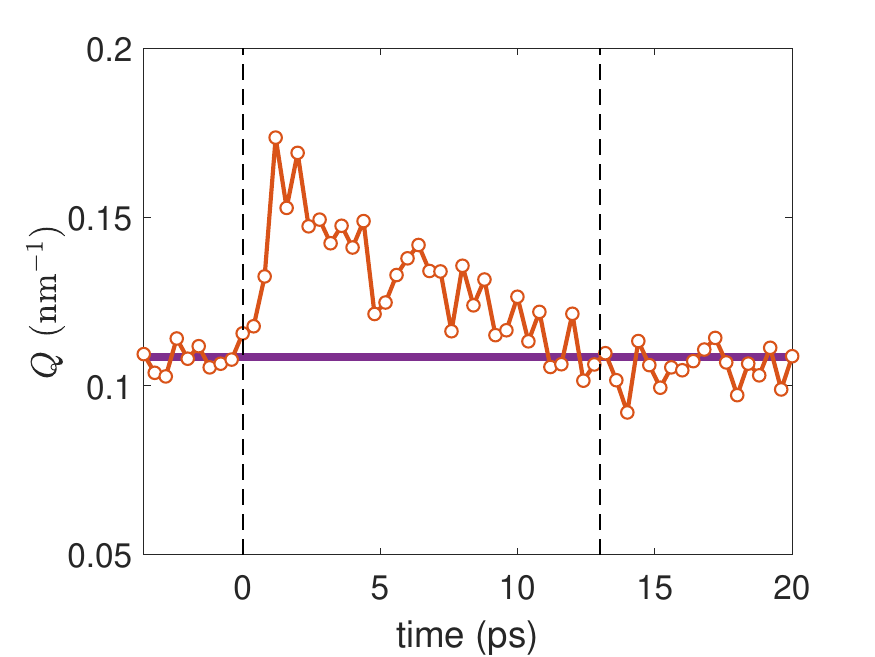}
\caption{ \label{fig:sup4} Time evolution of the fitted form factor, $Q(t)$. The magenta solid line indicates the pre-pump equilibrium value, $Q(0)$. The dashed vertical black lines indicate $t = 0$~ps and $t = 13$~ps, when $Q(t)$ recovers its pre-pump value. }
\end{figure}

We apply Eq.~\eqref{eq:2} to fit the time-dependent, azimuthally averaged scattering data. In Fig.~\ref{fig3}(a), (b), and (c), we show the fitting results in the form of $S_e(q,t)$ at select times. In all panels, the diffraction amplitudes are quenched, c.f. Fig.~\ref{fig2}(b), as expected for ultrafast demagnetization. The full temporal evolution of the normalized amplitudes $M_{1,3,5}(t)/M_{1,3,5}(t=0)$ is shown in Fig.~\ref{fig3}(d), exhibiting three distinct dependencies on time. At 1.6~ps, the third-order ring is quenched slightly more than the first-order ring. Both the first-order and third-order rings partially recover until 13~ps after quenching. For $t > 13$~ps, the third-order ring resumes quenching, but at a much slower rate of $\approx2$\% per picosecond. The fifth-order ring is still detectable in spite of a greatly reduced signal-to-noise ratio due to the low photon flux at high $q$. The error of the fifth-order ring amplitude shown as a gray background in Fig.~\ref{fig3}(d) is estimated from the fitted amplitude's fluctuations at $t<0$. It is still apparent that the 5th order ring amplitude is almost entirely quenched after pumping, despite the reduced signal-to-noise. By averaging the scattering data over a time-span from 6~ps to 11~ps, we are able to fit the fifth-order ring with better accuracy and confirm that its amplitude is quenched by almost 90 percent, see Appendix~\ref{app_b}. The almost total quench of the 5th order ring is important for the quantitative analysis presented below.

The time evolution of the form-factor, $Q(t)$, is shown in Fig.~\ref{fig:sup4}, exhibiting an ultrafast increase and subsequent recovery to equilibrium, shown by a solid magenta line, at $\approx10$~ps. It is possible this is the result of an ultrafast alteration in the characteristic exchange length of the sample. If it is indeed the case that $Q\propto1⁄\lambda_\mathrm{ex}$, then the ultrafast change in $Q$ would suggest that the exchange stiffness is attenuated more than the magnetization immediately after optical pumping. This is in agreement with previous studies that found significant evidence for a reduction in the exchange splitting in ultrafast pumping experiments~\cite{Turgut_2013,Turgut_2016}. The fact that $Q(t)$ returns to its equilibrium value $13$~ps after pumping suggests that this is the time scale at which conventional equilibrium concepts relating temperature, magnetization, and the renormalization of exchange, i.e., $A\propto M_s$, are valid~\cite{Bloch1962,Lowde1965,Stringfellow1968}. Coincidentally, $10$~ps is the time scale at which the electron, spin, and lattice thermal baths are generally considered to be in thermal equilibrium with each other.

\begin{figure}[t]
\centering \includegraphics[width=2.8in]{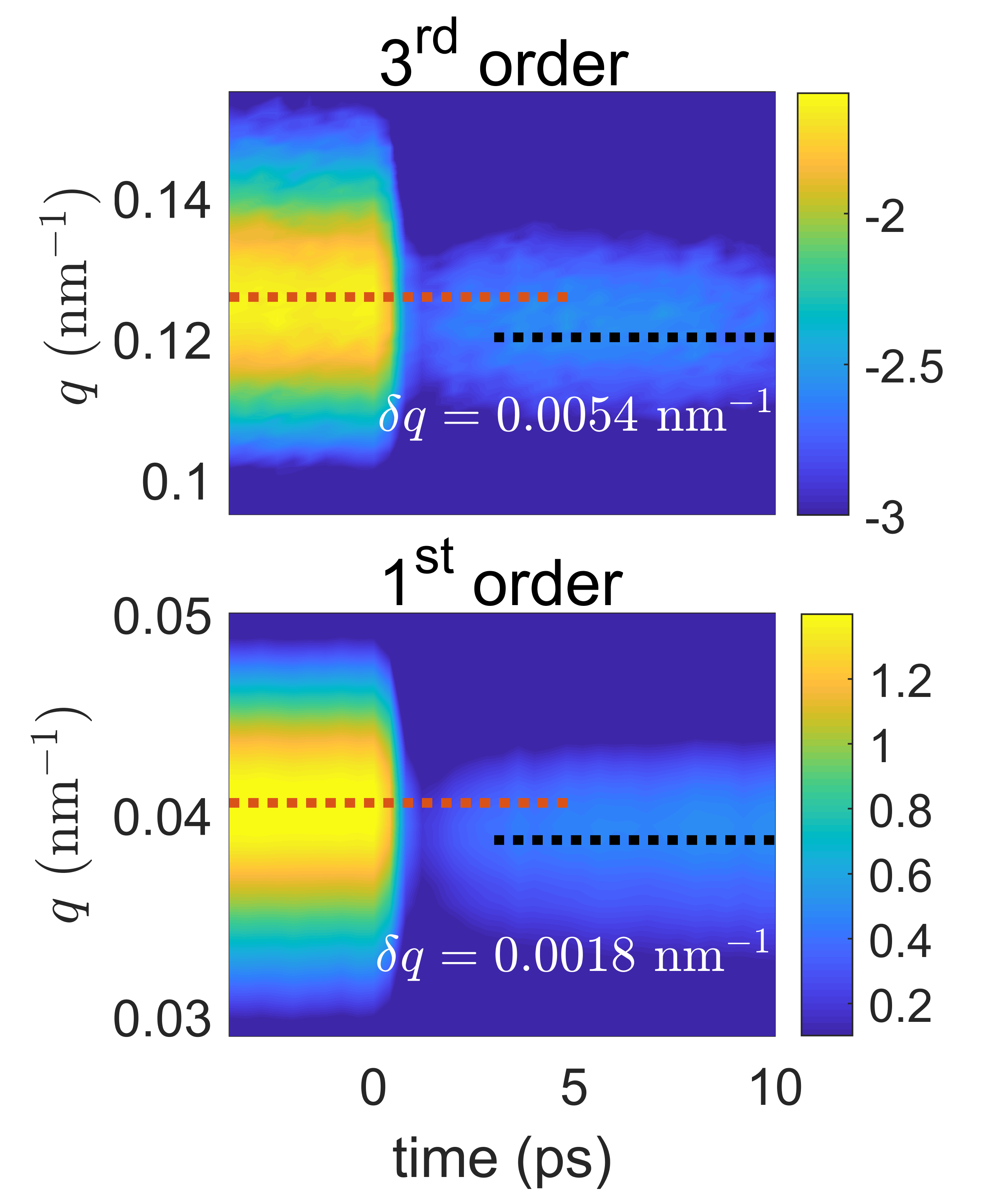}
\caption{ \label{fig4} Color contour plots of the azimuthally-averaged magnetic diffraction ring intensity profiles, after form-factor normalization, as a function of both time and radial $q$ for the first and third-order rings. The first and third-order rings are presented in the bottom and top panel, respectively. The dashed red line marks the average ring radius prior to optical pump. The dashed black line marks the ring radius averaged between $6$~ps and $1$~ps after optical pump. A shift in both rings is detected; $0.0018$~nm$^{-1}\pm0.0001$~nm$^{-1}$ for the first-order ring and $0.0054$~nm$^{-1}\pm0.0003$~nm$^{-1}$ for the third-order ring. }
\end{figure}

The diffraction ring radii also exhibit ultrafast changes. In Fig.~\ref{fig4} we show colour contour plots of the azimuthally averaged magnetic diffraction ring intensity profiles for the first- and third-order rings as a function of both time and $q$, divided by the equilibrium form factor. The visible shift in both radii are marked with horizontal lines that indicate the $q$-value for the time-averaged ring radii before (dashed red line) and between $6$ and $11$~ps after optical pumping (dashed black line). The difference in the average radii, $\delta q$, is $0.0018$~nm$^{-1}\pm0.0001$~nm$^{-1}$ for the first-order ring, and $0.0054$~nm$^{-1}\pm0.0003$~nm$^{-1}$ for the third-order ring. These differences in the average radii are harmonically related by a factor $3$, indicating that the full spectrum shifts.

A reminiscent 4\% shift in the first-order diffraction ring radius was previously observed by time-resolved x-ray scattering in a similar magnetic system that supports labyrinth domain patterns~\cite{Pfau2012}. In that work, the authors attributed the first-order diffraction ring radius shift to a Gaussian filter function that attenuated high-$q$ components of the diffraction after pumping. Use of a Gaussian filter function was justified as higher order effect to be expected under the hypothesis of strong ultrafast domain-wall broadening that fully quenches all higher order diffraction rings. However, higher order rings were not detected in Ref.~\onlinecite{Pfau2012}, precluding the ability to directly test the validity of this hypothesis. The substantial dynamic range of our experimental method allows us to directly access the necessary data to determine how pumping actually affects the domain wall width.

\section{Effective Bloch domain-wall model}

For materials with strong perpendicular magnetic anisotropy, a hyperbolic Bloch-wall model is applicable~\cite{Hubert2009}, with a one-dimensional (1D) profile given by
\begin{equation}
\label{eq:7}
    m_d(x,t)=m(t)\tanh\left(\frac{x}{a(t)}\right),
\end{equation}
where $m(t)$ is the time-dependent, normalized magnetization within the adjacent domains and $a(t)$ is a measure of the domain-wall width. Equation~\eqref{eq:7} is strictly applicable to materials with negligible second-order anisotropy constant. In our case, the ratio between the second and first-order anisotropy constants is $\kappa = -0.36$. This ratio leads to a broader domain-wall, yet similar in shape to that predicted from Eq.~\eqref{eq:7}. See, e.g., Figure 3.60 in Ref.~\onlinecite{Hubert2009}.

Our use of a particular equilibrium model for the domain wall profile is meant to be applied to our data analysis in the most general sense. In particular, it is powerful in its ability to provide a quantitative interpretation of the time-resolved diffraction intensities. However, the general intention of this model is to, at a minimum, provide a qualitative description of how the domain walls behave under conditions of ultrafast pumping. Any model for the domain wall profile will have a monotonically decreasing intensity of the diffraction rings with diffraction order. While the sharpness of the domain wall is somehow encoded in the dependence of intensity on $q$, we can only speculate as to the exact details of that wall profile since we can only measure those intensities up to 5th order. We indeed concede that the fact that our model actually fits our data so well, as we will show, does not necessarily mean the model is correct in an exact sense, but the ability to interpret the diffraction intensities in a quantitative manner does add confidence to any qualitative interpretation of the data that involves time-dependence of the domain wall profile.

The domain-wall width is calculated following the Lilley interpretation that considers the slope of the domain-wall profile at the origin\cite{Hubert2009}. Therefore, we define the domain-wall width as
\begin{equation}
\label{eq:8}
    w_w=\pi a(t).
\end{equation}

To extract the parameter $a(t)$ from the experimental data, one must be conscious that Eq~\eqref{eq:7} is a 1D profile whereas the sample is stabilized in a 3D labyrinth domain pattern. By micromagnetic simulations using mumax3~\cite{Vansteenkiste2014}, we find that the domains have negligible variation through the thickness (see Appendix~\ref{app_d}) and so the labyrinth domain pattern can be assumed to be 2D. In order to map this pattern to a 1D equivalent, we must consider that the labyrinth domain network distributes the diffracted photons uniformly along the azimuthal coordinate. To account for this, we consider that the domain-wall is part of a periodic array of domains of width  $w_d=\pi⁄q_0$ and the effective amplitudes $A_{n}(q_0,t)$ are related to the fitted diffraction ring amplitudes via
\begin{equation}
\label{eq:9}
    A_{n}(q_0,t) = M_{n}(t)\sqrt{2\pi nq_0}.
\end{equation}
This relationship means that we map the 2D diffraction data to the diffraction expected from a perfectly periodic 1D stripe domain pattern, which is the underlying assumption implied in azimuthal integration of scattered intensities and any derived analysis. We stress that, while we initially perform numerical azimuthal \textit{averaging} of the diffraction signal to determine the time-dependent scattering amplitude of the $n$th order ring, e.g. $M_{n}(t)$, we subsequently perform fits with the square root of the azimuthally \textit{integrated} intensity, e.g. $A_{n}(q_0,t)$. This is equivalent to the statement that the diffraction from the labyrinth domain pattern, when azimuthally integrated, is equivalent to the diffraction from 1D periodic domains, albeit with the diffracted photons spread out uniformly in the azimuthal coordinate. This allows us to connect the simplified 1D model to the actual 2D diffraction pattern so that our analysis is consistent throughout. 
\begin{figure}[t]
\centering \includegraphics[width=3in]{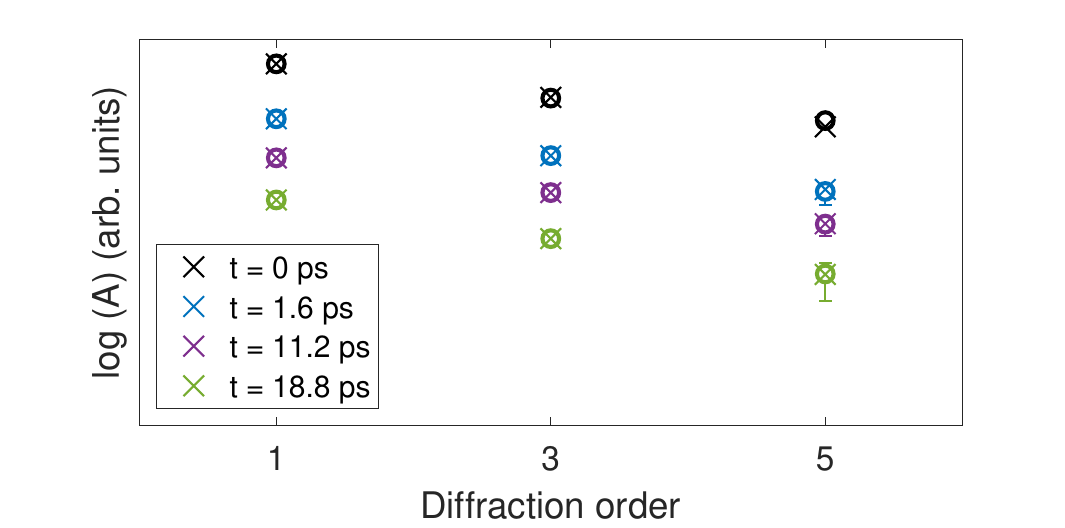}
\caption{ \label{fig5} Effective amplitudes at selected time instances fitted with the Bloch-wall model of Eq.~\eqref{eq:10}. The amplitudes (circles) and fits (crosses) are shown in logarithmic scale and vertically shifted for clarity. }
\end{figure}

To connect the effective amplitudes to the Bloch-wall model, we compute the Fourier transform of Eq.~\eqref{eq:7} by convolving the spectrum of a square wave of periodicity $w_d=\pi/q_0$ with the spectrum of the derivative of Eq.~\eqref{eq:7}. The resulting discrete spectrum has harmonic amplitudes given by
\begin{equation}
\label{eq:10}
    A_{n}(t)=\frac{\pi m(t)w_w(t)}{2w_d(t)}\mathrm{csch}\left(\frac{\pi nw_w(t)}{2w_d(t)}\right),
\end{equation}
where we used the domain-wall-width definition of Eq.~\eqref{eq:8} to explicitly include it in the expression. We restate that $m(t)$ is the asymptotic magnetization amplitude in the infinite wavelength limit, i.e. when $q_0\rightarrow0$. It is not to be confused with the maximum amplitude of the magnetization between the domains for non-zero $q_0$.

Fits to the effective amplitudes are shown in logarithmic scale by crosses in Fig.~\ref{fig5} for selected times. The amplitudes obtained from Eq.~\eqref{eq:9} and the fitted amplitudes of the azimuthally averaged scattering are shown by circles with errorbars denoting the standard deviation of the fit. We note that the fifth-order ring amplitude has little weight on the overall fit shown in Fig.~\ref{fig5} since its amplitude after pumping is close to the noise level. However, the nearly total quench of the fifth-order ring is consistent with the quantitative degree of domain-wall broadening extracted from the 1st and 3rd order diffraction rings.

\begin{figure}[t]
\centering \includegraphics[width=3in]{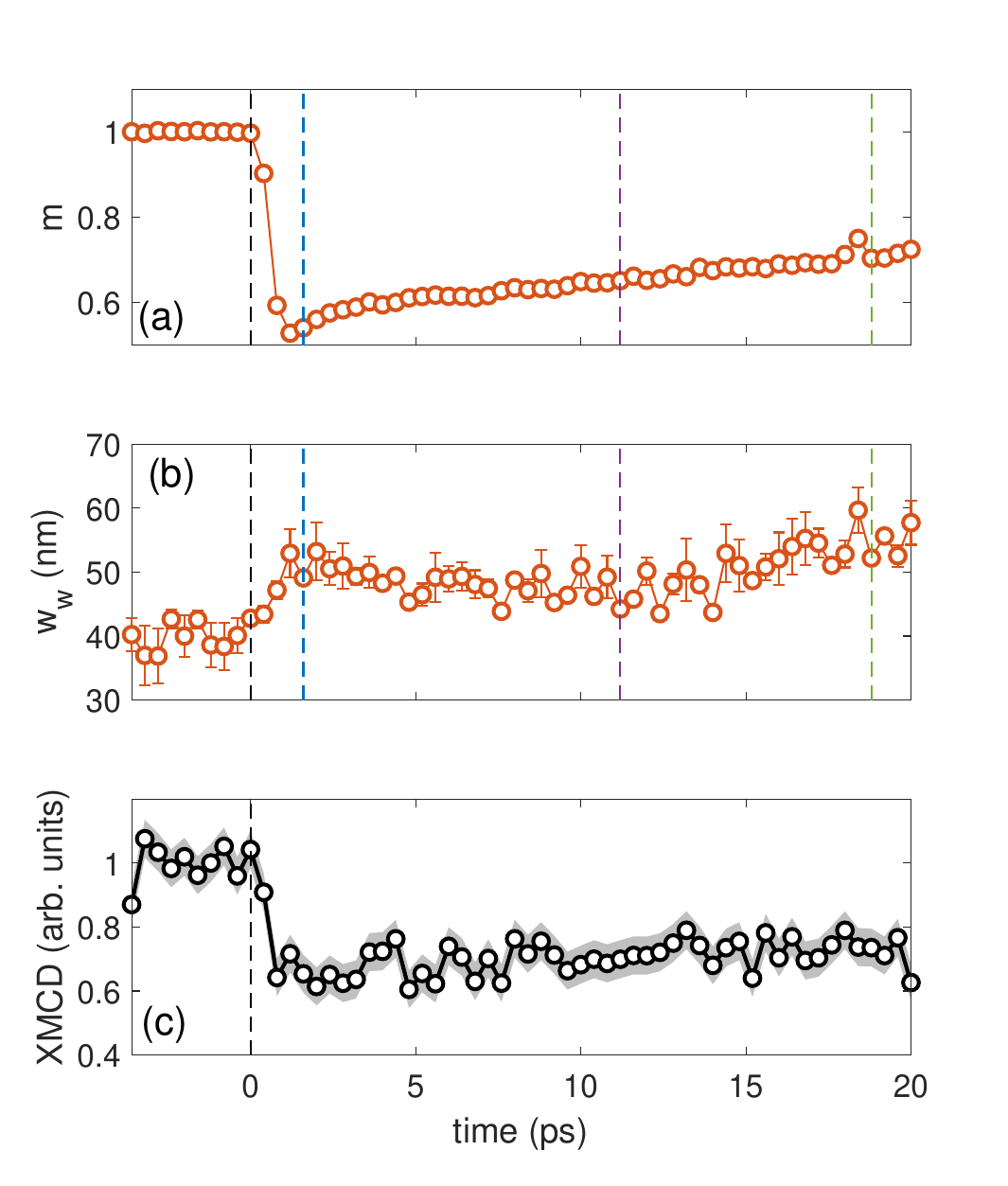}
\caption{ \label{fig5new} Evolution of the (a) asymptotic magnetization, $m(t)$, and (b) the domain-wall width, $w_w(t)$. The color-coded vertical lines represent the time instances shown in Fig.~\ref{fig5}. The domain walls broaden by $31$\% within the first $2$~ps after pumping. (c) XMCD data obtained when the sample is magnetized with the external field. The error is shown as gray shadow and it is computed from the standard deviation of the pre-pump data. }
\end{figure}

The evolution of $m(t)$ is shown in Fig.~\ref{fig5new}(a), exhibiting a typical demagnetization behavior, but with a faster remagnetization process than would otherwise be surmised by inspection of the diffraction amplitude data in Fig.~\ref{fig3}(d). It is also distinct from the time-resolved XMCD data, shown in Fig.~\ref{fig5}(c). While the signal-to-noise for the XMCD is significantly less than that of the diffraction data fitting, it is clear that there is little to no recovery of the magnetization after pumping, as also apparent in the diffraction amplitude data.

The evolution of $w_w(t)$ is shown in Fig.~\ref{fig5new}(b). The initial domain-wall width is $39$~nm, in good agreement with the calculated value of $45$~nm from Bloch-wall theory when considering the reduced anisotropy $\pi\sqrt{A_\mathrm{ex}⁄(K_1+K_2)}$. We find a significant broadening of the domain-walls from $39$~nm to $51$~nm ($31$\%) within $1.6$~ps, followed by partial recovery towards its original equilibrium value in the first $13$~ps after pumping. From $13$~ps to $20$~ps, the domain-walls resume broadening by approximately $38$\% more than the original equilibrium value, likely because of a reduction in the effective magnetic anisotropy of the sample due to delayed thermal diffusion through the sample thickness, see Appendix~\ref{app_e}.

Ultrafast domain-wall broadening was indeed previously inferred from 1st order diffracting ring measurements~\cite{Pfau2012}. However, the method used in Ref.~\onlinecite{Pfau2012} was purely inferential insofar as it was not possible to truly determine the smoothing of the domain walls from the width of the diffraction rings. As shown here, we precisely determine the broadening from an entirely different perspective: we take advantage of the substantial dynamic range of our measurement system, then apply a Fourier series decomposition method that relies on the simple fact that the domain-wall profile is actually encoded in the relative amplitudes of the diffraction orders. In other words, we do not infer the domain-wall widths, but rather we directly measure them.

\section{Linear filter function analysis}

We have conclusively shown that domain wall broadening does indeed occur in our system, whereby we rigorously measured the degree of broadening by use of Fourier analysis. Having settled the question as to whether ultrafast broadening actually does occur, we can now test whether use of a linear filter function to extract wall broadening from the time-resolved distortions of a \textit{single diffraction ring}, as was done in Ref. \onlinecite{Pfau2012}, is consistent with our rigorous quantitative result.

Let us consider an azimuthally averaged spectrum that has a simple functional shape, $f_t(q)$ given by the product of a Lorentzian diffraction lineshape and a Gaussian filter function
\begin{equation}
\label{eq:appB1}
    f_t(q) = \left[\frac{M(t)}{\left(\frac{q-q_0}{\Gamma(t)}\right)^2+1}\right]e^{-q^2/2\sigma(t)^2},
\end{equation}
where $M(t)$, $q_0$, and $\Gamma(t)$ are the Lorentzian's amplitude, peak position at equilibrium ($t<0$), and linewidth, respectively, and $\sigma(t)$ is the Gaussian's standard deviation. This Gaussian filter function in reciprocal space is used to model how the domain structure that gives rise to the Lorentzian lineshape might be smoothed by convolution with the same Gaussian, but in the form of its inverse transform in real space.

The maximum of Eq.~\eqref{eq:appB1} can be analytically computed from $\partial f_t(q)/\partial q=0$. Introducing the peak shift $\Delta q(t)=q_0-q_\mathrm{max}(t)$ as a function of the measured peak position $q_\mathrm{max}(t)$ and solving for $\sigma$, we obtain
\begin{equation}
\label{eq:appB2}
    2\sigma(t)^2=\frac{\Gamma(t)^2q_0+q_0\Delta q(t)^2-\Delta q(t)^3}{\Delta q(t)}-\Gamma(t)^2.
\end{equation}

For a small shift, $\Delta q(t) \ll 1$, we can approximate the Gaussian standard deviation to
\begin{equation}
\label{eq:appB3}
    2\sigma(t)^2\approx\Gamma(t)^2\left(\frac{q_0}{\Delta q(t)}-1\right).
\end{equation}

\begin{figure}[t]
\centering \includegraphics[width=3.2in]{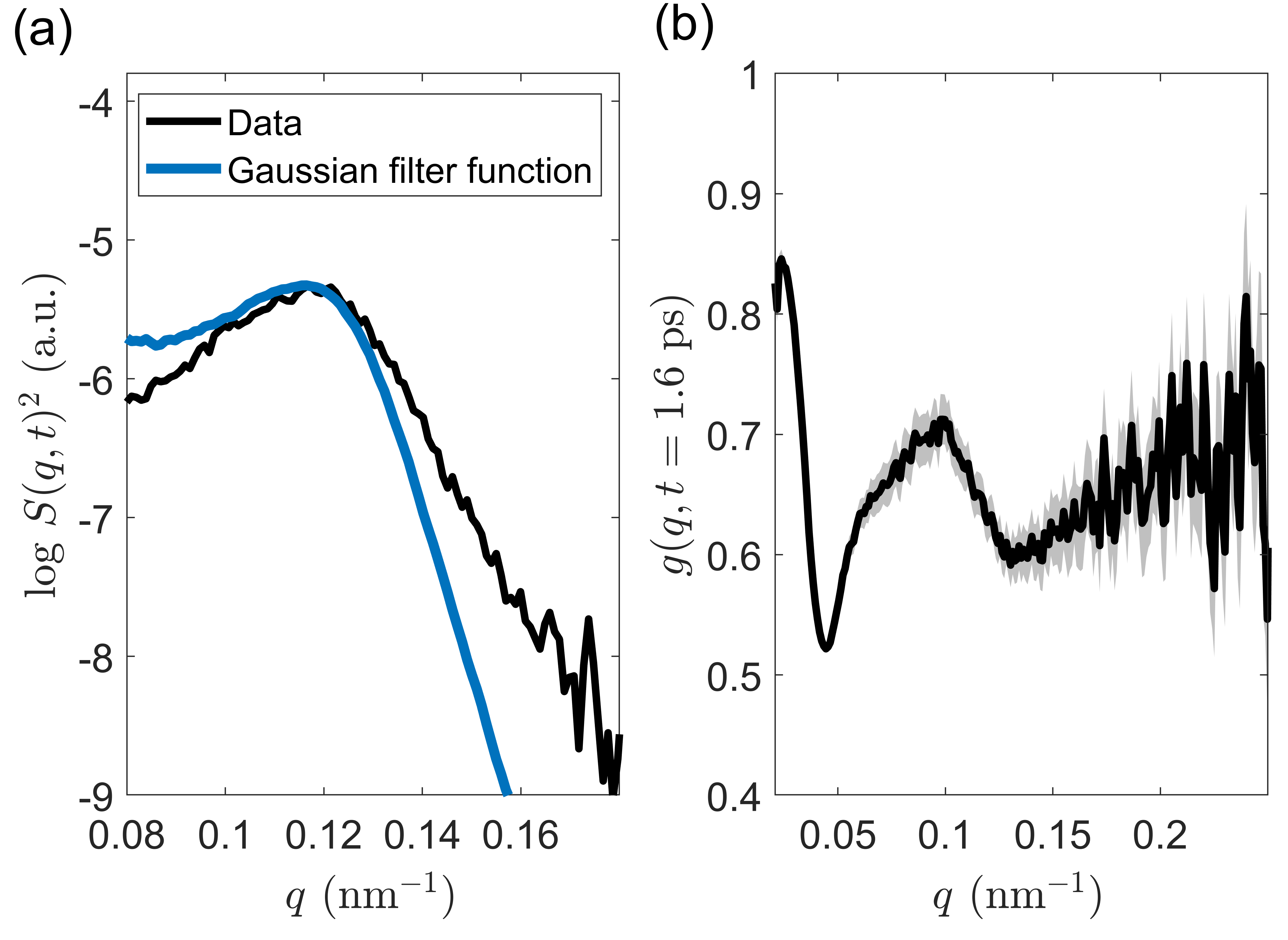}
\caption{ \label{fig7} (a) Comparison of the third-order ring spectrum isolated from experimental data (black curve) and predicted by the use of a suitable Gaussian filter function (blue curve). (b) Time-dependent filter function $g(q,t)$ computed from experimental data at $t = 1.6$~ps. The grey shaded area represents the error in determining $g(q,t)$ from shot noise. }
\end{figure}

This powerful yet simple expression permits us to directly determine the degree of smoothing $\sigma(t)$ as a function of the measured parameters, i.e. the original diffraction ring radius $q_0$, the shift in the ring radius $\Delta q(t)$, and the original width of the diffraction ring $\Gamma(t)$. For example, given the equilibrium ring radius $q_0\approx0.0392$~nm$^{-1}$ and parameters at $1.6$~ps after optical pumping $\Delta q\approx2.7$~$\mu$m$^{-1}$, and $\Gamma\approx8.8$~$\mu$m$^{-1}$, we obtain $\sigma\approx0.023$~nm$^{-1}$. \textit{This is a substantial degree of smoothing that implies an almost total quench of all diffraction orders except for the 1st order ring.} In the case of our measured distortions in the 1st order diffraction ring, the amplitude ratio between the third and first-order rings would be expected to be $\approx8\times10^{-6}$, \textit{four orders of magnitude smaller} than the actual fitted ratio of $0.08$. Clearly, the degree of domain wall broadening extracted from the distortions of the 1st order diffraction ring is inconsistent with that directly measured by use of the 3rd and 5th order diffraction rings. We are forced to conclude that the ultrafast shift of the diffraction ring radii cannot be ascribed to domain-wall broadening.

In addition to being quantitatively erroneous with regard to domain-wall broadening, the use of a filter function as a model for such broadening also induces a sizeable asymmetry in the $q$-dependent spectrum that is not actually observed. In Fig.~\ref{fig7}(a), we compare the isolated azimuthally averaged third-order ring from the experimental data (black curve) and computed by means of a Gaussian filter function (blue curve). For this, we use the pre-pump ($t < 0$) time-averaged data for the third-order ring, multiply it by a Gaussian filter function with a standard deviation computed from Eq.~\eqref{eq:appB3}, and scale it so as to match the ring amplitude at all times. As is easily seen by eye, the Gaussian filter results in significant asymmetry in the spectrum that does not agree with the experimental data. An even poorer agreement is obtained by scaling the Gaussian filter function by the measured XMCD data, as would be expected from the fact that a filter function at $q=0$ should be proportional to the quenching for a uniformly magnetized sample. 

A slightly different approach can be taken by assuming distortions of the domain configurations. In this case, a real-space distortion would imply a harmonic application of the Gaussian filter function, as further discussed in Appendix~\ref{app_c}. This approach also results in a poor model for our experimental data and is yet further evidence that fitting of the data with a Gaussian filter is not an effective method for the extraction of domain wall broadening from the measured diffraction.

It can be argued that the assumption of a Gaussian smoothing function to model domain-wall broadening is itself specious, and that a phenomenological smoothing function would be more accurate at assessing the details of domain wall dynamics. Based on linear response theory, the dynamic evolution of the domain network can always be analyzed in terms of a time-dependent spatial filter kernel $G(x,y,t)$ that is convolved with the equilibrium perpendicular-to-plane magnetization component
\begin{equation}
\label{eq:4}
    M_z(x,y,t) = G(x,y,t)*M_z (x,y,t=0).
\end{equation}

Because the scattering intensity is related to $M_z$ via a Fourier transform, $|S(\mathbf{q},t)|^2=|\mathcal{F}\{M_z \}|^2$, it is possible to reinterpret the filtering kernel as a multiplicative factor in Fourier space, $g(\mathbf{q},t)$, that describes the time-dependent evolution of the scattering, given by
\begin{equation}
\label{eq:5}
    g(\mathbf{q},t)=\sqrt{\frac{|S(\mathbf{q},t)|^2}{|S(\mathbf{q},t=0)|^2}},
\end{equation}
where $g(\mathbf{q},t)=\mathcal{F}\{G(x,y,t)\}$. This kernel may be computed from experimental data with good accuracy up to the third-order peak. The associated error to the kernel can be computed by standard error propagation to be
\begin{equation}
\label{eq:6}
    \delta g(\mathbf{q},t)=\frac{1}{\sqrt{N(\mathbf{q}}}\frac{g(\mathbf{q},t)}{2}\sqrt{\frac{1}{q(\mathbf{q},t)^2}+1},
\end{equation}
where $N(\mathbf{q})$ is the time-independent photon count per $\mathbf{q}$ and we assume that the main source of noise in the measurement is shot noise.

The resultant scalar filter function obtained from experimental data at $t = 1.6$~ps is shown in Fig.~\ref{fig7}(b). The shaded area represents the error computed from Eq.~\eqref{eq:6}. We see that the spectral modulation of the diffraction rings by such a time-dependent scalar filter function is qualitatively non-trivial, given the function's complex oscillatory form. We also see that the reciprocal-space radii of maximal quenching are correlated with the diffraction ring positions. If optically-induced quenching is a simple function of wavelength, whereby shorter wavelength features demagnetize more efficiently than longer wavelengths, we would expect monotonically decreasing low-pass filter behavior. Instead, we find that the experimentally determined filter function is unambiguously non-monotonic. Such a complex dependence of quenching cannot be easily attributed to an uncomplicated reciprocal-space dependence of demagnetization, as was originally proposed by Pfau, \textit{et al.}\onlinecite{Pfau2012}.

Finally, the analysis presented in this section also clarifies the impact that thermal fluctuations of the domain wall spatial positions has on diffraction. When thermal fluctuations affect atomic coordinates in the case of Bragg diffraction, this is typically accounted for with the Gaussian-like Debye-Waller factor (DWF). The DWF accounts for the fact that the amplitudes of high-order diffraction features ~\cite{Ruhle1996} are strongly attenuated by thermal fluctuations when $q$ is greater than the wavenumber of the scattering photons. If we apply by analogy the DWF to the case of magnetic diffraction from domain wall patterns, and we assume thermal fluctuations cause random fluctuations of domains walls from their average position, the estimated $\sigma\approx0.023$~nm$^{-1}$ from Eq.~\eqref{eq:appB3} would imply a mean spatial displacement of $\sqrt{2\pi/(2\sigma^2)}\approx77\times10^{-9}$~m. In other words, domain walls would need to fluctuate approximately $77$~nm to account for the observed shift in the ring positions. Such massive spatial fluctuations in the domain wall positions are clearly unphysical. In addition, such large random domain wall motion would also fully extinguish the higher order scattering from the domain structure by smearing out details of the domain wall structure, which is clearly not the case in the measurements presented here. As such, we can also rule out any stochastic phenomena that can be accounted for with the DWF as relevant explanations for the phenomena we observed.

\section{Ultrafast domain rearrangement}

\begin{figure}[t]
\centering \includegraphics[width=3in, trim={0 0 0 0.5in}, clip]{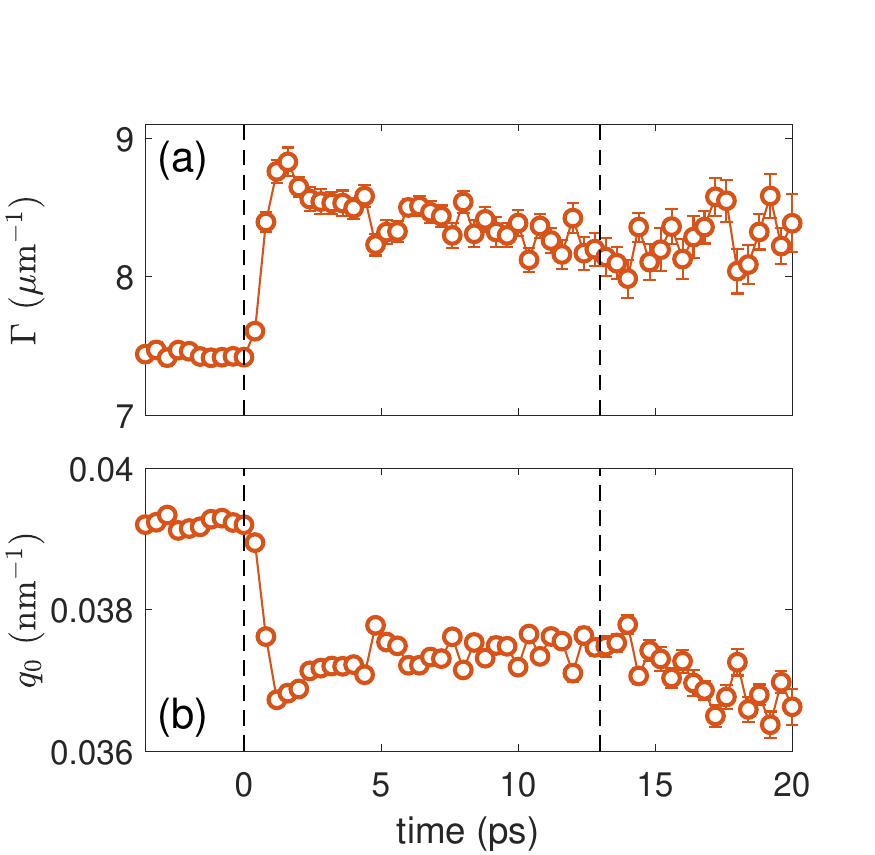}
\caption{ \label{fig6} Time evolution of the fitted (a) linewidth $\Gamma$ and (b) ring radius $q_0$. Both quantities exhibit ultrafast changes. The ring radius shrinks by 6\% and the linewidth broadens by 15\%. }
\end{figure}

Having established and definitively quantified ultrafast domain-wall broadening from the time dependence of diffraction ring amplitudes, and excluding the applicability of a monotonically decaying filter function as the result of said wall broadening, we now focus on the ultrafast dynamics manifest in both the diffraction ring radii and linewidths. The diffraction rings’ linewidths exhibit an ultrafast time dependence, as shown in Fig.~\ref{fig6}(a). The linewidth broadens 15\% within $1.6$~ps and exhibit a partial recovery until $13$~ps where it then remains relatively constant. Interpreting the linewidth $\Gamma$ as a reciprocal measure of the correlation length, $2\pi⁄\Gamma$, this quantity decreases from the equilibrium value of $845$~nm to $711$~nm at $1.6$~ps after optical pumping. We understand such correlation lengths to be the distances over which the phase of the periodic domain structure is no longer predictable. As such, a rapid reduction in the correlation length is a clear indication that the domain pattern is subject to some sort of ultrafast rearrangement, whereby the phase coherence of the domains is further reduced under far-from-equilibrium conditions. We stress that the fits are obtained by the use of Lorentzian functions that are symmetric with respect to the ring radii. As such, the small error bars in Fig.~\ref{fig6}(a) are clear evidence that there is minimal ultrafast distortion of the ring shapes beyond position and width.

The decrease in the spatial coherence of the domain structure implies that some degree of domain-wall motion is occurring, albeit such motion must be both stochastic and zero-mean. The need for the motion to be stochastic is obvious given the broadening of the linewidth. The requirement that the domain wall motion be a zero-mean effect is necessary if there is no net overall change in the average domain wall density. Instead, we conjecture that ultrafast optical pumping simply causes increased variability in the size of the domains across the probe spot. Because the scattering is fitted well with a symmetric Lorentzian lineshape, we can rely on standard Fourier analysis of a Langevin-type equation subject to a Wiener random process~\cite{Gardiner2004} to quantify a domain-wall “jitter” $\delta\lambda(t)$ given by
\begin{equation}
\label{eq:11}
    \delta\lambda(t) = \sqrt{\frac{2\Gamma}{q_0^3}},
\end{equation}
that depends on the measured linewidth and ring radius. Using Eq.~\eqref{eq:11}, we obtain $\delta\lambda(t) = 15.7$~nm~$\pm0.1$~nm at equilibrium. $\delta\lambda(t)$ jumps to $18.8$~nm~$\pm0.2$~nm in a time $1.6$~ps after pumping. The average domain-wall speed required to accommodate such a rapid change in positional jitter is approximately $2$~km/s, well within the range of plausibility.~\cite{Balaz2020}

The evolution of ring radii is shown in Fig.~\ref{fig6}(b). We detect a 6\% shift in both the first- and third-order ring radii at $1.6$~ps after optical pumping, followed by a partial recovery. This confirms the similar shift from labyrinth domain patterns observed in Refs.~\onlinecite{Pfau2012} and \onlinecite{ZhouHagstrom2022}. Here, we extend our observation to the higher-order diffraction rings. After $13$~ps, the rings continue to shrink for the remainder of the measurement time. The very slight continued shrinking of the ring radius shift past $13$~ps might be attributed at least in part to thermal expansion of the lattice.

The simultaneous observation of both a reduction of the correlation length and a contraction of the diffraction rings’ radii are strong evidence in support for the spatial rearrangement of the domain pattern at picosecond timescales. These results indicate that domain-walls are \textit{mobile} upon optical excitation, though the exact details of the spatial modifications are not directly accessible via scalar diffraction measurements nor by the required averaging of single-shot measurements.

The hypothesis of domain dilation stemming from the ring radius shift was deemed unphysical in Ref.~\onlinecite{Pfau2012} due to the exceedingly large domain-wall speeds at the edges of the x-ray probe spot implied by a fractional expansion in the average domain width. In our experiments, a 6\% domain dilation would imply domain-wall speeds at the outer radius of our probe spot to be on the order of $(0.06\times10\mu\mathrm{m})⁄1.6$~ps~$\approx375$~km/s. Such an extreme speed is many orders of magnitude faster than what is capable by any known mechanisms to drive isolated domain-walls in equilibrium, e.g. Refs.~\onlinecite{Wang2012,Thiaville2005}, and even one order of magnitude faster than a recent prediction of femtosecond domain-wall motion due to superdiffusive spin currents~\cite{Balaz2020}. As such, the physical interpretation of the observed contraction in the ring radii remains a mystery. Nevertheless, it still stands that we have confirmed that domain wall broadening does in fact occur, as Pfau, et al.,~\cite{Pfau2012} concluded, but we disproved their method of determining the magnitude of the broadening. In addition, we have now shown that the effect extends harmonically to the third-order diffraction ring, thereby verifying that the effect is both  reproducible and self-consistent across multiple orders of diffraction. This clearly precludes any kind of experiment artifact as the origin of the broadening and shifts.

\section{Conclusion}

In this paper, we have measured the time-dependent scattering of a labyrinth domain pattern subject to ultrafast optical pumping up to the fifth-order diffraction ring. This large range of wavevectors allowed us to use Fourier series concepts to unambiguously and simultaneously recover information on the average domain-wall width and domain size. Our main observation is that, contrary to what has been posited to date, domain-walls are mobile when subjected to ultrafast optical pumping, suggesting the appearance of a net torque on the domain-walls and the ultrafast rearrangement of the domain pattern.

Our results are consistent with the recent observation that a much larger shift in the radii of the diffraction rings occurs in the case of labyrinth domains as opposed to that for parallel stripe domains~\cite{ZhouHagstrom2022}. This indicates that lower symmetry spin textures, e.g. labyrinths with randomly dispersed domain junctions, termini, and bubbles, are more susceptible to ultrafast rearrangement than higher symmetry spin textures, e.g. periodic stripe domains.

A recent study has indicated that the nature of the torque may be related to the hybridization of domain-wall types in materials with Dzyaloshinskii-Moriya interaction~\cite{Leveille2020}. It would be interesting to generalize this concept to domain-walls in material without well-defined chirality.

The possibility of domain structure rearrangement by ultrafast optical pumping raises the possibility that spin textures can be optically controlled. The presented results greatly expand the parameter space in which to further explore the rich nature of far-from-equilibrium magnetization dynamics, including materials with more exotic spin textures, such as chiral domain networks and skyrmion lattices.

\begin{acknowledgments}
This material is based upon work partially supported by the U.S. Department of Energy, Office of Science, Office of Basic Energy Sciences under the x-ray Scattering Program Awards No. DE-SC0002002 and DE-SC0018237 and Award Number DE-SC0017643. Operation of LCLS is supported by the U.S. Department of Energy, Office of Basic Energy Sciences under contract No. DE-AC02-76SF00515. M.A.H. was partially supported by NSF CAREER DMS-1255422. The authors thank S. Zohar for his support with the data collection and analysis.
\end{acknowledgments}

\appendix

\section{Subtraction of charge contribution to the scattering intensity}
\label{app_a}

To separate the electronic and magnetic contributions, we use the refractive index formalism to describe the interaction of soft x-rays with the ferromagnetic multilayer film. An equivalent description in terms of scattering amplitudes is also possible~\cite{Tesch2013}.

Within the approach adopted here, spatial variations of the refractive index $n_0$ will cause the incident x-rays to scatter. The variations could either be caused by inhomogeneities of the chemical composition or surface roughness, collectively referred to as charge variations $c(\mathbf{r})$, where $\mathbf{r}=(x,y)$ is the spatial coordinate, or a spatially dependent profile of the out-of-plane magnetization component $s(\mathbf{r})\equiv M_z(\mathbf{r})$. We assume no spin-charge correlation in our multilayered samples. While such correlations exist for granular\cite{Granitzka2017,Reid2018} or patterned~\cite{Liu2015b} media, we do not expect such correlations in our samples because the exchange interaction between adjacent grains is comparable to that within the grains themselves. Indeed, such strong intergranular exchange coupling is a prerequisite for the formation of labyrinth domain patterns.

With the corrections to the refractive index $\delta n_c$ and $\delta n_s$ due to charge and spin variations, respectively, the electric field of an electromagnetic wave transmitted through the sample is
\begin{equation}
\label{eq:appA1}
    E=E_0e^{ikd[n_0+\delta n_ss(\mathbf{r})+\delta _cc(\mathbf{r})]},
\end{equation}
where $E_0$ is the incident circularly polarized wave and is assumed to be a plane wave $(E_0=1)$ due to the large spot size of the incident beam of $\approx100$~$\mu$m relative to its wavelength of $1.45$~nm, $d$ is the sample thickness, and $k$ is the wavenumber of the incident x-rays. $E$ in Eq.~\eqref{eq:appA1} is referred to as the exit surface wave (ESW). We divide out the term $e^{ikdn_0}$ and, to make the notation more compact, introduce substitutions $C(\mathbf{r})=ikd\delta n_cc(\mathbf{r})$ and $S(\mathbf{r})=ikd\delta n_ss(\mathbf{r})$. A Taylor expansion of Eq.~\eqref{eq:appA1} to first order in $C(\mathbf{r})$ and $S(\mathbf{r})$ yields
\begin{equation}
\label{eq:appA2}
    E=1+C(\mathbf{r})+S(\mathbf{r}).
\end{equation}

The scattered intensity at the detector is obtained by taking a Fourier transform of Eq.~\eqref{eq:appA2} and multiplying it by the conjugate
\begin{equation}
\label{eq:appA3}
    I(\mathbf{q}) = |C(\mathbf{q})|^2+|S(\mathbf{q})|^2+2\mathrm{Re}\{C(\mathbf{q})S(\mathbf{q})\},
\end{equation}
where $C(\mathbf{q})$ and $S(\mathbf{q})$ are Fourier transforms of $C(\mathbf{r})$ and $S(\mathbf{r})$, respectively. The Fourier transform of the first term in Eq.~\eqref{eq:appA2} is a delta function $\delta(\mathbf{q})$, which is non-zero only when the scattering vector $q=0$. Since we are not interested in the unscattered signal, we neglected the delta function in Eq.~\eqref{eq:appA3}.

Because the incident x-ray probe is circularly polarized, the magnetically and electronically scattered x-rays have the same polarization, and thus the third term in Eq.~\eqref{eq:appA3} is, in general, non-zero.
\begin{figure}[t]
\centering \includegraphics[width=3in]{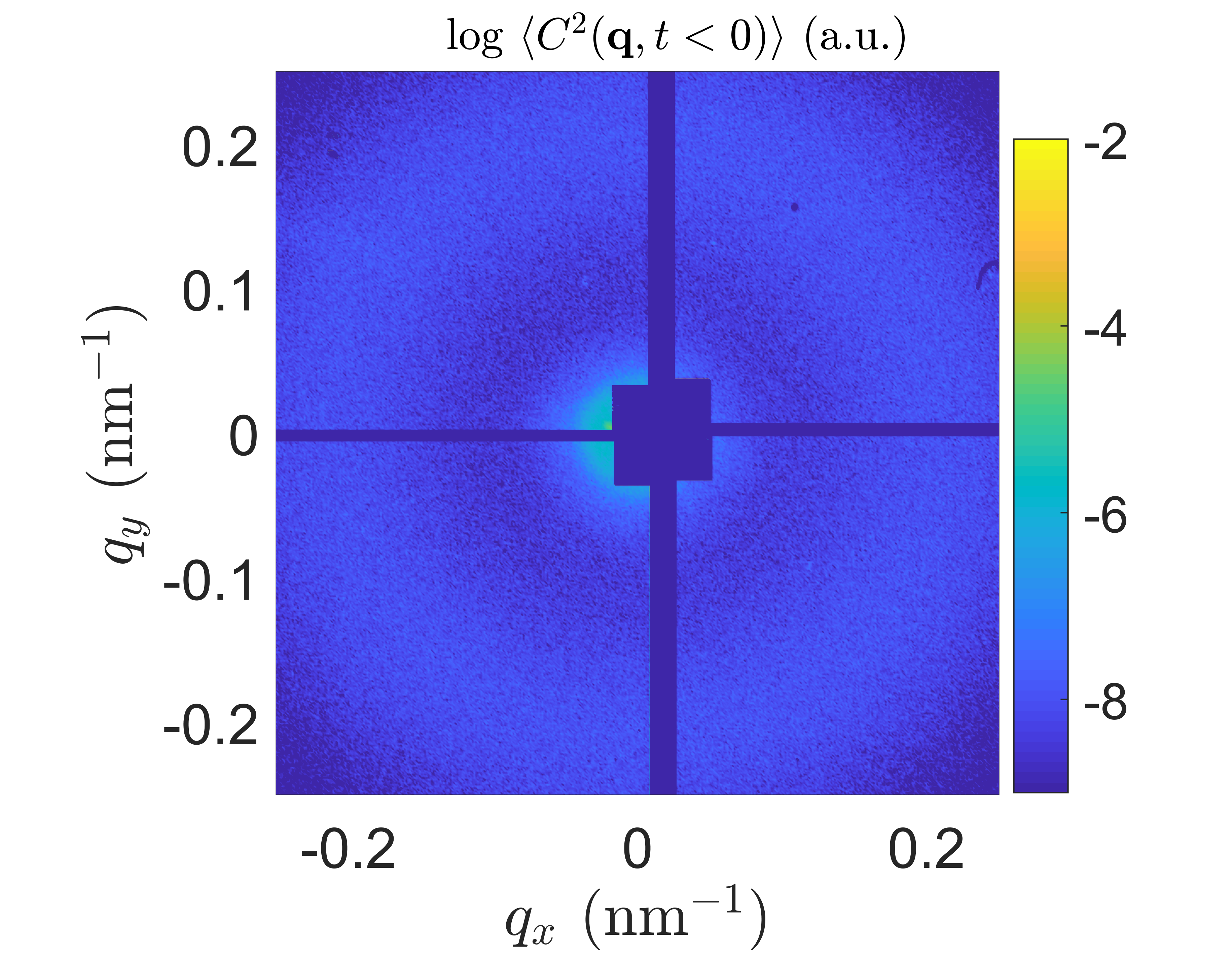}
\caption{ \label{figAppA} Pre-pump ($t<0$) charge scattering. }
\end{figure}

When a saturating perpendicular magnetic field $H_z$ is applied to the sample, it eliminates the magnetic domains, and the complex magnetically scattered signal $S(\mathbf{q})$ vanishes except at $q=0$, in which case $I(\mathbf{q})\propto|C(\mathbf{q})|^2$. However, the total transmission through the sample still depends on its magnetization direction due to the effect of x-ray magnetic circular dichroism upon circularly polarized x-rays, which in turn affects the charge scattering because of the non-zero sample thickness. Thus, the magnitude of the scattered intensity is essentially a product of the charge scattering and a field-dependent XMCD transmission factor $I_\mathrm{XMCD}(H_z)|C(\mathbf{q})|^2$. This variation can be accounted for by including second order terms in the Taylor expansion of Eq.~\eqref{eq:appA2}, as was done in Ref.~\onlinecite{Zusin2018_thesis}. For that reason, the pure charge scattering $|C(\mathbf{q})|^2$ with circular polarized x-rays and a non-negligible sample thickness is found from the scattering intensities taken with positive and negative applied saturating fields
\begin{equation}
\label{eq:appA4}
    \Sigma = \frac{1}{2}\left[I(\mathbf{q},+H_z)+I(\mathbf{q},-H_z)\right]=|C(\mathbf{q})|^2.
\end{equation}

The charge scattering shown in Fig.~\ref{figAppA} is found at wavevnumbers of $0.2$~nm$^{-1}$, so that the signal mainly overlaps with the magnetic fifth-order diffraction ring and the effect of cross-terms in Eq.~\eqref{eq:appA3} is negligibly small on the magnetic first-order and third-order diffraction rings. Further details can be found in Ref.~\onlinecite{Zusin2018_thesis}. We then extract the magnetic scattering intensity as
\begin{equation}
\label{eq:appA5}
    |S(\mathbf{q})|^2 = I(\mathbf{q},H_z=0)-\Sigma.
\end{equation}

\section{Fits of time averaged scattering data after pumping}
\label{app_b}
\begin{figure}[b]
\centering \includegraphics[width=3in]{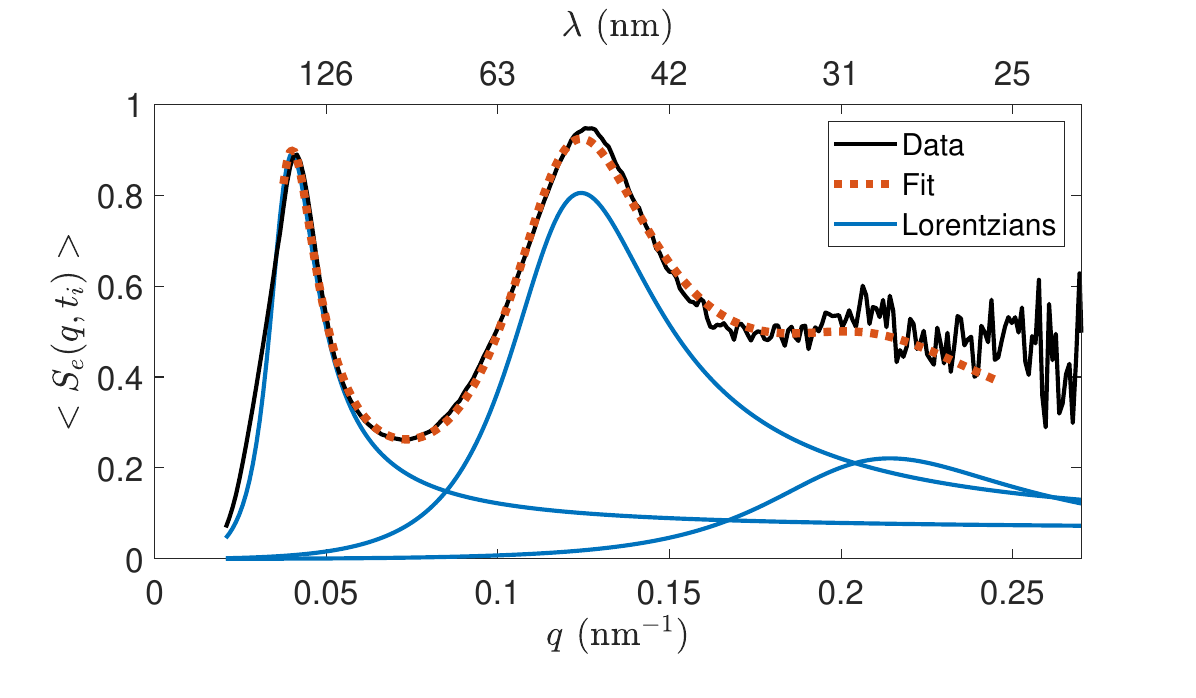}
\caption{ \label{figAppB} Fits of time averaged scattering data after pumping. }
\end{figure}

To test if the fifth-order ring can be fitted accurately from the data, we time-average the already azimuthally averaged scattering in the interval $t_i =$~[$6$~ps, $11$~ps]. The equalized data, fits, and Lorentzians in the representation of Eq.~\eqref{eq:3} are shown in Fig.~\ref{figAppB} by solid black, dashed red, and solid blue curves, respectively. After time-averaging, the fifth-order ring, though significantly quenched relative to what is detected prior to pumping, e.g. Fig.~\ref{fig2}(a), is more clearly distinguished, and is found to be resolved in a manner that is consistent with the fitting of the first- and third-order rings. Small errors in the fit with increasing $q$ are visibly enhanced in this equalized representation. Regardless, the fit is extremely sensitive to the exact positions of the diffraction rings in the data, as captured with the Lorentzian model for the ring profiles. In this particular fit, we obtained $q_0 = 0.0373 $~nm$^{-1}\pm4.5\times10^{-5}$~nm$^{-1}$. This represents a $\approx0.26$\% shift in the ring radius of $0.0374$~nm$^{-1}\pm9\times10^{-5}$~nm$^{-1}$ calculated from time-average of the domain width shown in Fig.~\ref{fig5}(a) within the time interval $t_i$.

\section{Modeling the Ultrafast Distortion of the Domain Configuration}
\label{app_c}
\begin{figure}[t]
\centering \includegraphics[width=2.5in]{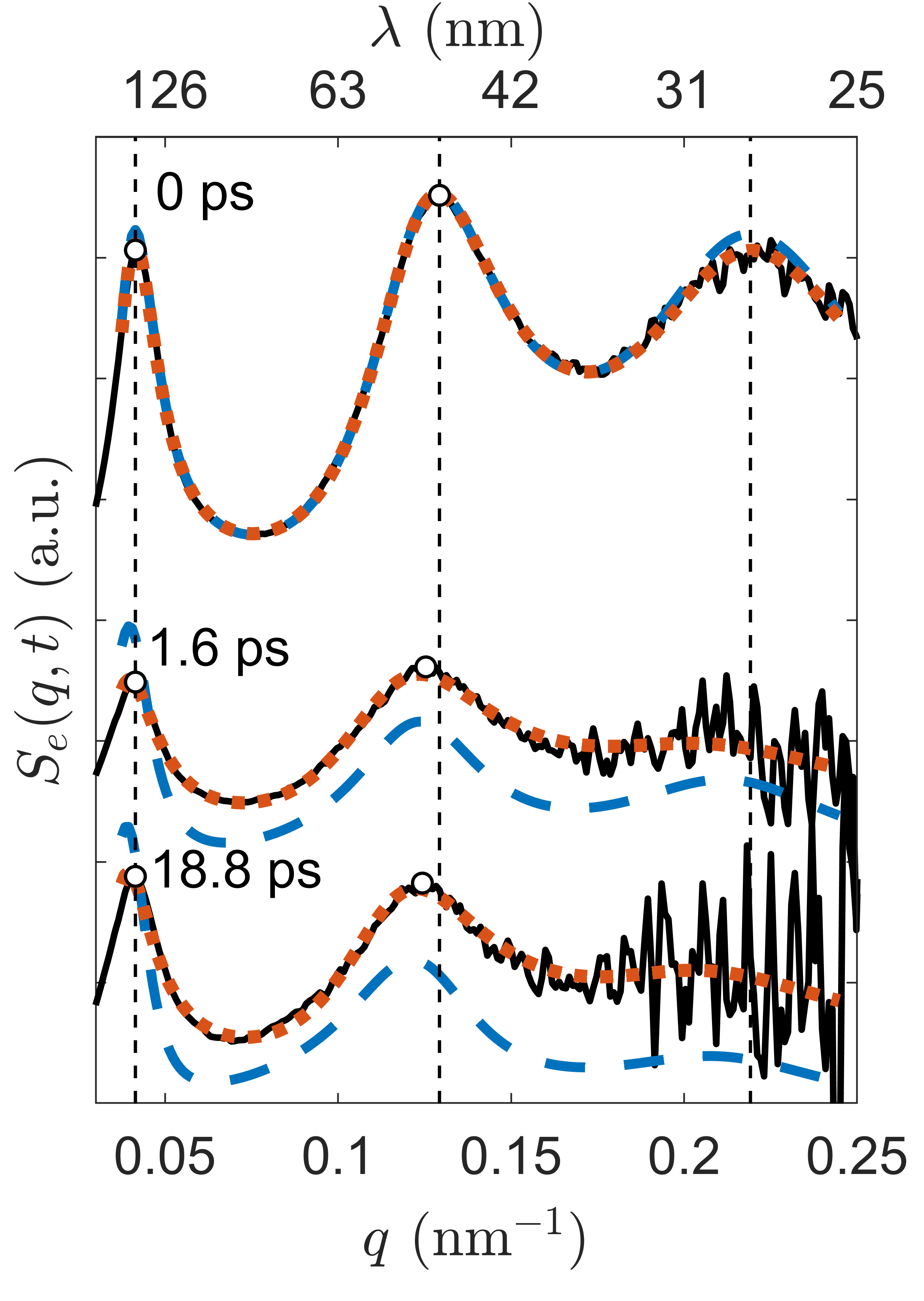}
\caption{ \label{FigAppC} Data fitting using a fitted Gaussian filter function. The data is presented in the equalized representation of Eq.~\eqref{eq:2}. The experimental data are shown by solid black curves, the empirical fits by dashed red curves, and the fits using a Gaussian filter function, Eq.~\eqref{eq:appC1} by dashed blue curves. Time instances at $0$~ps, $1.6$~ps, and $18.8$~ps are shown and vertically shifted for clarity. }
\end{figure}
\begin{figure*}[t]
\centering \includegraphics[width=6in]{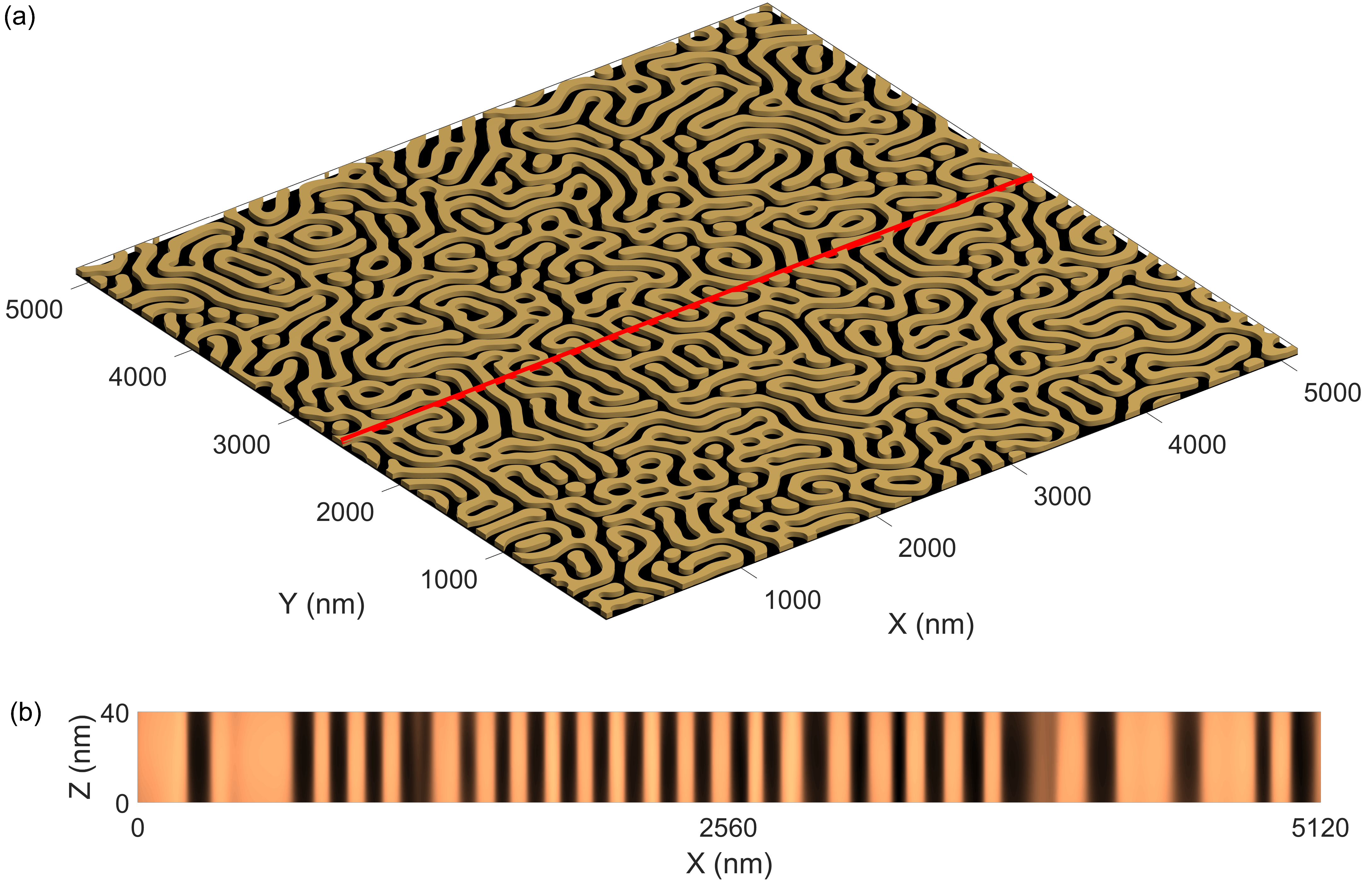}
\caption{ \label{figAppD} (a) 3D rendering of the equilibrium magnetization. (b) Cross-section of the domain pattern along the red line shown in (a) and the thin film’s height. The profile is uniform across the thickness. }
\end{figure*}

An alternative interpretation of our data is that there is an ultrafast distortion in the shape of the individual diffraction rings due to a change in the statistical distribution of domain sizes, shapes, and/or amplitudes. In that case, the distortion in the spectral profile of each diffraction ring is a function of the ring order. We consider a simplified model for such an effect whereby the distortion in the ring profile can be represented by a Gaussian filter function that shifts the weight of each diffraction ring to lower wavenumbers. To fit the experimental data with such a model, we multiply each Lorentzian in the diffraction ring profile by a filter function that depends on the diffraction order of the ring to be fitted, i.e. the position and width of the filter function are linear in the diffraction order $n$. Multiplication of each diffraction ring by a different filter function is justified based on the fact that the domain-wall profile gives rise to the multiple observed rings according to Fourier series decomposition. The resulting empirical function $f_G(q,t)$ has the following functional form
\begin{widetext}
\begin{equation}
\label{eq:appC1}
    f_G(q,t) = e^{-2q/Q(0)}\left[M_0(0)+\sum_{n=1,3,5}{\frac{\tilde{M}_{n}(t)}{\left(\frac{q-nq_0(0)}{n\Gamma(0)}\right)^2+1}e^{-q^2/2\left(n\sigma(t)\right)^2}}\right]^2
\end{equation}
\end{widetext}
where the quantities $\tilde{M}_{n}(t)$ are fitting parameters and $\sigma(t)$ is the Gaussian standard deviation. Note that only the Gaussian filter function is time-dependent while the remaining parameters can be estimated from the fit to the equilibrium spectrum.

The resulting fits at selected times are shown in Fig.~\ref{FigAppC}. At equilibrium, the Gaussian reduces to a constant and the scattering is accurately fitted as shown by the dashed blue curve. However, the scattering at later times cannot be fitted with this functional form as shown by the blue curves. For comparison, we also plot the fits using our empirical function in red dotted curves.

\section{Micromagnetic simulations}
\label{app_d}
\begin{figure*}[t]
\centering \includegraphics[width=6.5in]{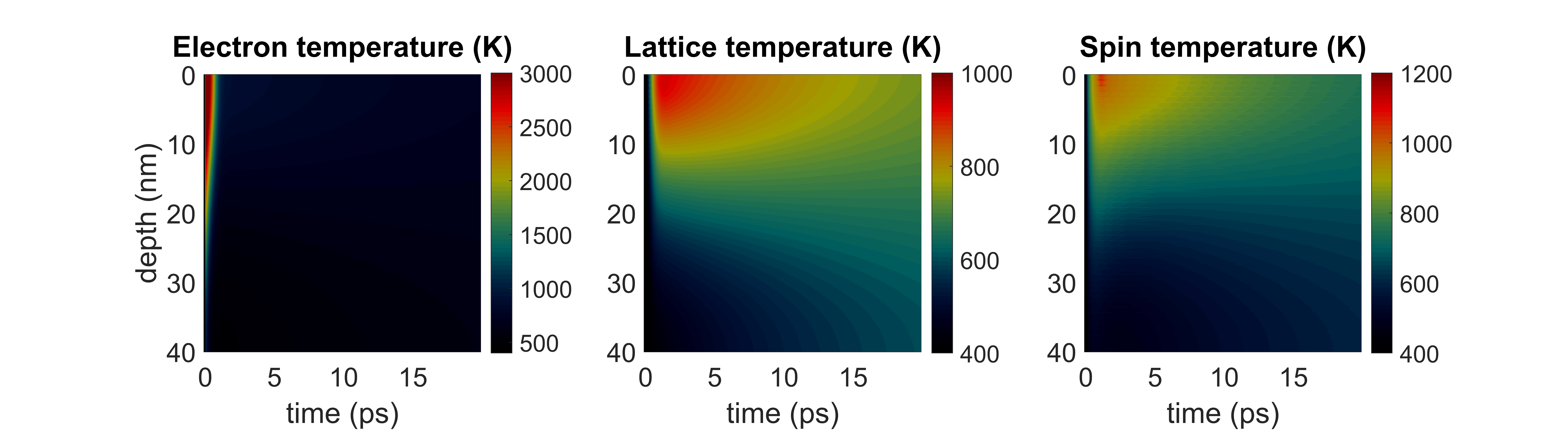}
\caption{ \label{figAppE} Depth profile temperature results of three-temperature heat transport model for the electronic, lattice, and spin temperatures. }
\end{figure*}
The equilibrium domain pattern is estimated using the micromagnetic package MuMax3~\cite{Vansteenkiste2014} and run on an NVIDIA Tesla P100 GPU accelerator. We set a $5,120$~nm$\times5,120$~nm$\times40$~nm simulation area discretized in cells of $5$~nm$\times5$~nm$\times5$~nm and imposing periodic boundary conditions on the film’s plane. The cell size is below the estimated exchange length of $7.3$~nm. We used experimentally measured magnetic parameters at room temperature: $M_s = 771$~kA/m, $K_1 = 739$~kJ/m$^3$, and $K_2 = -266$~kJ/m$^3$ and we assumed an exchange constant of $A_\mathrm{ex}=20$~pJ/m.

The ground state is found by the use of the \texttt{relax} routine in Mumax3 which estimates the energy minimum by removing the conservative term of the Landau-Lifshitz equation and using a Runge-Kutta 23 solver.

The simulation is initialized with a random magnetization distribution. The resulting labyrinth domain pattern is shown in Fig.~\ref{figAppD}(a). By Fourier analysis, we obtain an average domain size of $78$~nm~$\pm13$~nm, in good agreement with the equilibrium domain size deduced from the pre-pump scattering data. The cross section along the red line shows that the domains are essentially constant across the thickness, as shown in Fig.~\ref{figAppD}(b). 

\section{Time evolution of thermal profile in thick metallic multilayers after optical pumping}
\label{app_e}

The heat transport in the film was modeled by use of the three-temperature model~\cite{Beaurepaire1996}, which includes three coupled equations to describe the dynamics of the electron, lattice, and spin temperature baths
\begin{widetext}
\begin{subequations}
\begin{eqnarray}
    C_\mathrm{e}(T_\mathrm{e})\frac{\partial T_\mathrm{e}}{\partial t} &=& \vec{\nabla}\cdot\left(\kappa_\mathrm{e}(T_\mathrm{e},T_\mathrm{l})\vec{\nabla}T_\mathrm{e}\right)+G_\mathrm{el}(T_\mathrm{e})\left(T_\mathrm{l}-T_\mathrm{e}\right)+G_\mathrm{es}\left(T_\mathrm{s}-T_\mathrm{e}\right)+S(z,t),\\
    C_\mathrm{l}(T_\mathrm{l})\frac{\partial T_\mathrm{l}}{\partial t} &=& \vec{\nabla}\cdot\left(\kappa_\mathrm{l}(T_\mathrm{e},T_\mathrm{l})\vec{\nabla}T_\mathrm{l}\right)+G_\mathrm{el}(T_\mathrm{e})\left(T_\mathrm{e}-T_\mathrm{l}\right)+G_\mathrm{ls}\left(T_\mathrm{s}-T_\mathrm{l}\right),\\
    C_\mathrm{s}(T_\mathrm{s})\frac{\partial T_\mathrm{s}}{\partial t} &=& G_\mathrm{es}\left(T_\mathrm{e}-T_\mathrm{s}\right)+G_\mathrm{ls}\left(T_\mathrm{l}-T_\mathrm{s}\right).
\end{eqnarray}
\end{subequations}
\end{widetext}

We used material-specific and temperature-dependent values for the specific heat $C_x$, thermal conductivity $\kappa_x$, the electron-lattice coupling constant $G_\mathrm{el}$, the electron-spin coupling constant $G_\mathrm{es}$, and the lattice-spin coupling constant $G_\mathrm{ls}$~\cite{Zusin2018_thesis,Anisimov1997,Ivanov2003,Lin2008}. The subscript $x$ stands for e, l, or s to denote the electron, lattice, or spin system, respectively. The laminate structure of the sample was taken into account, and the spatial profile of the heat source $S(z,t)$ was found by computing the absorption of the pump light with an incident fluence of $26.7$~J/cm$^2$ by the film using the multilayer formalism of Ref.~\onlinecite{Zak1991}. More details on the material parameters used in the simulation can be found in Ref.~\onlinecite{Zusin2018_thesis}.

The calculated depth-dependent electron, lattice, and spin temperatures are shown in Fig.~\ref{figAppE}. The magnetization profile was obtained from the calculated temperature of the spin system using the experimentally measured temperature dependence of the magnetization. The electron-spin coupling parameter was chosen to be $G_\mathrm{es}=3\times10^{17}$~W/m$^3$K to obtain a good fit to the experimental XMCD signal. However, one must take the calculated temperatures for the various thermal baths in this model to be no more than rough estimates at the short times over which substantial changes in the magnetic scattering occurs. Given that $G_\mathrm{es}\gg G_\mathrm{ls}$, and an estimated $C_\mathrm{s}(T_\mathrm{s})\approx1.5\times10^6$~J/m$^3$K at the elevated temperatures expected after pumping, the estimated time constant for heat transfer between the electronic and spin system is $5$~ps, which is much longer than the measured domain dilation time and the domain-wall broadening time of $1.6$~ps. This highlights the fact that the electron-spin scattering processes in the far-from-equilibrium regime are strongly amplified for this system when compared to those expected from highly simplified models based on equilibrium dynamics.

\end{document}